\documentclass[prd,twocolumn,superscriptaddress,floatfix,amsmath,amssymb,amsfonts,nofootinbib,longbibliography]{revtex4-2}

\usepackage{multirow}
\usepackage{float} 
\usepackage{scalerel}
\usepackage[normalem]{ulem}
\usepackage[english]{babel}
\usepackage{graphicx}
\usepackage{dcolumn}
\usepackage{bm}
\usepackage{blindtext}
\usepackage{verbatim}
\usepackage{relsize}
\usepackage{mathrsfs}
\usepackage{musicography}
\usepackage{amsmath}
\usepackage{blindtext}
\usepackage{cancel}
\usepackage{physics}
\usepackage{epstopdf}
\usepackage{mathtools}
\usepackage{blindtext}
\usepackage{tensor}
\usepackage{color}
\usepackage[usenames,dvipsnames]{pstricks}
\usepackage{epsfig}
\usepackage{pst-grad} 
\usepackage{pst-plot} 
\usepackage{hyperref}
\usepackage{verbatim}
\usepackage{slashed}
\usepackage{dsfont}

\usepackage{cellspace}
\newcommand{\mf}{\mathsf}

\newcommand{\ii}{\mathrm{i}}

\renewcommand{\L}{\mathcal{L}}

\newcommand{\tc}[1]{\textsc{#1}}

\newcommand{\tbb}[1]{\textcolor{black}{#1}}

\allowdisplaybreaks[1] 

\begin{document}

\title{The role of quantum degrees of freedom of relativistic fields\\in quantum information protocols}



\author{T. Rick Perche}
\email{trickperche@perimeterinstitute.ca}

\affiliation{Department of Applied Mathematics, University of Waterloo, Waterloo, Ontario, N2L 3G1, Canada}
\affiliation{Perimeter Institute for Theoretical Physics, Waterloo, Ontario, N2L 2Y5, Canada}
\affiliation{Institute for Quantum Computing, University of Waterloo, Waterloo, Ontario, N2L 3G1, Canada}

\author{Eduardo Mart\'in-Mart\'inez}
\email{emartinmartinez@uwaterloo.ca}

\affiliation{Department of Applied Mathematics, University of Waterloo, Waterloo, Ontario, N2L 3G1, Canada}
\affiliation{Perimeter Institute for Theoretical Physics, Waterloo, Ontario, N2L 2Y5, Canada}
\affiliation{Institute for Quantum Computing, University of Waterloo, Waterloo, Ontario, N2L 3G1, Canada}

\begin{abstract}
    We analyze the differences between relativistic fields with or without quantum degrees of freedom in relativistic quantum information protocols. We classify the regimes where the existence of quantum degrees of freedom is necessary to explain the phenomenology of interacting quantum systems. We also identify the precise regimes where quantum fields can be well approximated by quantum-controlled classical fields in relativistic quantum information protocols. Our results can be useful to discern which features are fundamentally different in classical and quantum field theory.
\end{abstract}

\maketitle

\section{Introduction}

All interactions which can be derived from the Standard Model rely on quantum fields as mediators. However, there are many regimes in which some of the fundamental properties of these mediating quantum fields are not relevant. In these cases, one can treat the fields as non-dynamical, establishing effective potentials that set a direct interaction between quantum systems.
{The question of whether a particular physical phenomenon is fundamentally due to the quantum nature of a field requires one to identify the relevant quantum features in interactions between quantum sources. This is particularly important if one wants to propose an experiment that can distinguish if a particular field has quantum degrees of freedom or not.}

This distinction is also especially relevant in the study of quantum information protocols implemented through quantum fields, as is usually considered in  the field of relativistic quantum information (RQI). Among its topics of study, RQI explores quantum information tasks that are implemented through local interactions with relativistic quantum fields. Examples of protocols that can be implemented via local operations in quantum field theory are entanglement harvesting~\cite{Valentini1991,Reznik1,reznik2,Salton:2014jaa,Pozas-Kerstjens:2015,HarvestingQueNemLouko,Pozas2016,HarvestingSuperposed,Henderson2019,bandlimitedHarv2020,ampEntBH2020,mutualInfoBH,threeHarvesting2022,twist2022}, entanglement farming~\cite{Farming}, quantum energy teleportation~\cite{teleportation,teleportation2014}, quantum collect calling~\cite{Jonsson2,collectCalling,PRLHyugens2015} and noise-assisted quantum communication~\cite{KojiCapacity,Ahmadzadegan2021}. 


The main goal of this manuscript is to understand the role played by the {quantum degrees of freedom of} a relativistic field in mediating interactions between quantum systems, paying special attention to relativistic quantum information protocols. We study communication and entangling protocols between two localized quantum systems, exploring the channel capacity of communication channels mediated by fields with or without quantum degrees of freedom, and quantifying the amount of entanglement that can be acquired by probes that interact via the field. We will also identify the regimes where the interaction of localized quantum systems with a quantum field can be well approximated by a simpler interaction with a {quantum-controlled} classical field (which has no quantum degrees of freedom of its own). 

In this paper we will show that there are indeed cases where fully featured quantum fields cannot be modelled by relativistic quantum-controlled classical fields with no \tbb{quantum} degrees of freedom. Specifically, these  cases are when the interactions are either short or strong, or when the sources of the field are not in \tbb{causal} contact with each other. On the other hand, we show that in the regimes where none of these three conditions is satisfied, it is not possible to distinguish whether a  field theory  has quantum degrees of freedom or not.

Establishing the regimes that can actually distinguish fields {with or without quantum degrees of freedom} can also be used to identify experimental settings which can witness whether a field is fundamentally quantum or not. For instance, this reasoning has been applied in the context of the gravitational field in~\cite{remi,boris}, where sufficiently precise experiments were argued to have the potential to witness the quantum nature of the gravitational field.

This manuscript is organized as follows. In Section \ref{sec:classFields} we briefly review how classical fields are coupled to classical sources, and how to write the field's dynamics in terms of the currents that source it. In Section \ref{sec:class} we use these results to model the {interaction between quantum sources via a quantum-controlled classical field with no quantum degrees of freedom}. In Section \ref{sec:quant} we describe the interaction of localized quantum sources with quantum fields. In Section \ref{sec:communication} we discuss the differences between {considering a field with or without quantum degrees of freedom} in communication protocols. In Section \ref{sec:entangling} we study {the role of the field's quantum degrees of freedom in} entangling protocols. In Section \ref{sec:entComparison} we compare the quantum-controlled classical field model with the quantum field theory model and identify the conditions {where the quantum degrees of freedom of the field can be neglected}. The conclusions of our work can be found in Section \ref{sec:conclusions}.

\section{Classical fields sourced by classical currents}\label{sec:classFields}

In this section we briefly review the interaction of two classical systems which source a classical field. Consider two general classical systems labelled by $\tc{A}$ and $\tc{B}$ in (3+1) Minkowski spacetime {(denoted by $\mathcal{M}$)} with Lagrangian densities $\mathcal{L}_\tc{a}$ and $\mathcal{L}_\tc{b}$, and an arbitrary field (scalar, spinor, vector, tensor, etc)  $\phi^a$ with Lagrangian $\mathcal L_\phi$. Here $a$ stands for any collection of Lorentz indices, corresponding to the spin of the field. The dynamics of the system will be prescribed by the action
\begin{equation}\label{eq:action}
    S = \!\int \!\dd V \!\left(\L_\textsc{a}\!+\!\L_\textsc{b}\! + \!\L_\phi \!-\! \lambda j^{(\textsc{a})}_a(\mf x) \phi^a(\mf x)\!-\! \lambda j^{(\textsc{b})}_a(\mf x) \phi^a(\mf x)\right),
\end{equation}
where $\dd V = \dd^4 \mf x$ is the Minkowski spacetime volume element, $j_a^{(\textsc{i})}(\mf x)$ are current densities associated with the systems labelled by $\textsc{I} \in \{\textsc{A},\textsc{B}\}$ and $\lambda$ is the coupling constant for the interaction. 

The equation of motion for the field is obtained by varying the action of Eq. \eqref{eq:action} with respect to the field:
\begin{equation}
    \pdv{\L_\phi}{{\phi^a}}  - \partial_{\mu}\left(\pdv{\L_\phi}{(\partial_\mu \phi^a)}\right) = \lambda j_a^{(\textsc{a})}(\mf x) + \lambda j_a^{(\textsc{b})}(\mf x).
\end{equation}
We will assume that the equation of motion above is linear, in the sense that it can be written as
\begin{equation}\label{eq:P}
    \mathcal{P}[\phi_a(\mf x)] = \lambda j_a^{(\textsc{a})}(\mf x) + \lambda  j_a^{(\textsc{b})}(\mf x),
\end{equation}
where $\mathcal{P}$ is a linear differential operator. We will further assume that the solution of the equation of motion $\mathcal{P}[\phi^a(\mf x)] = \lambda j^a(\mf x)$ can be written in terms of a retarded Green function $G_R^{ab}(\mf x,\mf x')$:
\begin{equation}
    \phi^a(\mf x) = \lambda \int \dd V G_R^{ab}(\mf x, \mf x') j_b(\mf x').
\end{equation}

We will work under the assumption that systems $\textsc{A}$ and $\textsc{B}$ are the only sources for the field. That is, if they were not present, the field would vanish.
Under this assumption for the field, and neglecting the self-energy of systems $\tc{A}$ and $\tc{B}$, one obtains the following Hamiltonian for the system in an inertial frame $(t,\bm x)$,
\begin{align}\label{eq:classHphi}
    H(t) = H_\textsc{a}(t) &+H_\textsc{b}(t) \\
    &+  \frac{\lambda}{2}\int \dd^3 \bm x\left(j^{(\textsc{a})}_a(\mf x) \phi_\textsc{b}^a(\mf x)+ j^{(\textsc{b})}_a(\mf x) \phi_\textsc{a}^a(\mf x)\right),\nonumber
\end{align}
where $H_\textsc{a}(t)$ and $H_\textsc{b}(t)$ are the free Hamiltonians for systems $\tc{A}$ and $\tc{B}$, and
\begin{equation}\label{eq:phiRet}
    \phi^a_\textsc{i}(\mf x) =  \lambda\int \dd V' G_R^{ab}(\mf x,\mf x') j_b^{(\textsc{i})}(\mf x')
\end{equation}
is the field sourced by system $\tc{I}$. Thus, the interaction of two classical systems with a field $\phi^a(\mf x)$ via currents $j^{(\textsc{i})}_a(\mf x)$ is associated with the interaction Hamiltonian
\begin{equation}
    H_\text{int}(t) =  \frac{\lambda}{2}\int \dd^3 \bm x\,\left(j^{(\textsc{a})}_a(\mf x) \phi^a_{\textsc{b}}(\mf x)+j^{(\textsc{b})}_a(\mf x) \phi^a_{\textsc{a}}(\mf x)\right),
\end{equation}
and using Eq. \eqref{eq:phiRet}, the Hamiltonian can be rewritten as
\begin{align}\label{eq:classHjj}
    H_\text{int}(t) =& \frac{\lambda^2}{2}\int \dd^3 \bm x\,\dd V' G_R^{ab}(\mf x,\mf x')\\
    &\times\left(j^{(\textsc{a})}_a(\mf x) j_b^{(\textsc{b})}(\mf x')+j^{(\textsc{b})}_a(\mf x)j_b^{(\textsc{a})}(\mf x')\right).\nonumber
\end{align}
For reference, we note that
\begin{align}
    &\int \dd t\, H_\text{int}(t) =\frac{\lambda^2}{2}\int \dd V \dd V' j^{(\textsc{a})}_a(\mf x)\Delta^{ab}(\mf x, \mf x')j_b^{(\textsc{b})}(\mf x'),\nonumber
\end{align}
where 
\begin{equation}\label{eq:deltaGRGA}
    \Delta^{ab}(\mf x, \mf x') = G_R^{ab}(\mf x,\mf x')+G_A^{ab}(\mf x,\mf x')
\end{equation}
and $G_A^{ab}(\mf x,\mf x')=G_R^{ba}(\mf x',\mf x)$ is the advanced propagator for the differential operator $\mathcal{P}$. The Hamiltonian of Eq. \eqref{eq:classHjj} then prescribes the dynamics of the classical systems $\tc{A}$ and $\tc{B}$ in terms the field's propagator, without requiring to specify the field's degrees of freedom.

\subsubsection*{Pointlike systems coupled to a real scalar field}

A simple but instructive example is to consider that the classical systems are pointlike and undergo fixed inertial trajectories $\mf z_\textsc{i}(t)$. Let us also consider that they are coupled to a massless real scalar field $\phi(\mf x)$ and have monopole moments $\mu_\tc{i}(t)$ which source the field. In this case the field satisfies the K.G. equation  with \mbox{$\mathcal{P} = \Box = \partial_\mu\partial^\mu$},
\begin{equation}
    \Box\phi(\mf x) = 0,
\end{equation}
and the retarded Green's function is
\begin{equation}
    G_R(\mf x,\mf x') = - \frac{1}{4\pi}\delta^{(4)}\left(-(t-t')^2+(\bm x-\bm x')^2\right)\theta(t-t'),
\end{equation}
where $\theta(t)$ denotes the Heaviside theta function.

The coupling of the systems with the field can be modelled by considering current densities \mbox{$j^{(\textsc{i})}(\mf x) = \mu_\textsc{i}(t) \delta^{(3)}(\bm x - \bm z_\textsc{i}(t))$}. In this case we obtain the Hamiltonian density
\begin{align}
    H_\text{int}(t) = \,\frac{\lambda^2}{2}\!\!\int \dd t'\!\big(&\mu_\textsc{a}(t)\mu_\textsc{b}(t')G_R(\mf z_\textsc{a}(t),\mf z_\textsc{b}(t'))\\&+\mu_\textsc{b}(t)\mu_\textsc{a}(t')G_R(\mf z_\textsc{b}(t),\mf z_\textsc{a}(t'))\big),\nonumber
\end{align}
so that
\begin{equation}
    \int \dd t\, H_\text{int}(t) = \frac{\lambda^2}{2}\int \dd t \dd t'\mu_\textsc{a}(t)\mu_\textsc{b}(t')\Delta(\mf z_\textsc{a}(t),\mf z_\textsc{b}(t')),
\end{equation}
Notice that for this example we are not considering the dynamics of the particles' trajectories (they are fixed). However, we have internal dynamics for the monopoles $\mu_\textsc{a}$ and $\mu_\textsc{b}$. 

\section{Quantum Systems Coupled via a classically propagated interaction}\label{sec:class}

In this section we describe how two quantum systems {can interact via a field} using the model presented in Section \ref{sec:classFields}. {In essence, the quantum systems source a field which propagates according to its classical equations of motion implemented by the retarded Green's function. This gives rise to what we will call a quantum-controlled field, (that we will also refer to as \textit{qc-field}), which has no quantum degrees of freedom of its own.} This will allow us to study examples where {qc-fields} can be used for quantum information protocols between two quantum systems and will allow us to compare them with protocols mediated by a fully featured quantum field. We start this section by studying two pointlike two-level systems coupled to a real scalar {qc-field}, and later generalize the formalism, allowing for smeared sources and more general spins.

\subsection{Pointlike two-level systems coupled to a real scalar field}\label{sub:classPointlike}

In order to consider quantum sources for the model analyzed at the end of Section \ref{sec:classFields}, we consider the case where the monopoles that source the field, $\mu_\textsc{a}$ and $\mu_\textsc{b}$, are quantum. In order to implement this, we associate  a Hilbert space $\mathcal{H}_\tc{i}\cong  \mathbb{C}^2$ to each system, and consider the monopoles to be described by observables \mbox{$\hat{\mu}_\textsc{i}(0) = \hat{\sigma}^+_\textsc{i} + \hat{\sigma}^-_\textsc{i}$}, where $\hat{\sigma}^\pm_\tc{i}$ are the SU(2) ladder operators of system $\tc{I}$.

The free dynamics for the systems is implemented by the free Hamiltonians
\begin{equation}
    \hat{H}_\textsc{a} = \Omega \hat{\sigma}^+_\textsc{a}\hat{\sigma}^-_\textsc{a}, \quad 
    \hat{H}_\textsc{b} = \Omega \hat{\sigma}^+_\textsc{b}\hat{\sigma}^-_\textsc{b}.
\end{equation}
We picked $\hat{H}_\tc{i}$ noncommuting with $\hat{\mu}_\tc{i}(0)$ in order to implement non-trivial dynamics for the monopoles\footnote{Notice that, while simple, this model is the analogue to the Unruh-DeWitt detector model~\cite{Unruh1976,DeWitt} interacting with a classical field instead of a quantum one, as we will discuss in detail later.}. We further introduce switching functions $\chi_\textsc{a}(t)$ and $\chi_\textsc{b}(t)$, which control the time duration of the interactions, so that the interaction Hamiltonian for the particles in the interaction picture can be written as
\begin{align}\label{eq:classHpointlike}
    \hat{H}_\text{int}(t) \!= \!\frac{\lambda^2}{2}\!\!\int &\dd t' \Big(\chi_\textsc{a}(t)\chi_\textsc{b}(t')\hat{\mu}_\textsc{a}(t)\hat{\mu}_\textsc{b}(t')G_R(\mf z_\textsc{a}(t),\mf z_b(t'))\nonumber\\
    &\!+\chi_\textsc{b}(t)\chi_\textsc{a}(t')\hat{\mu}_\textsc{b}(t)\hat{\mu}_\textsc{a}(t')G_R(\mf z_\textsc{b}(t),\mf z_\textsc{a}(t'))\Big),
\end{align}
where the interaction picture monopole moments are $\hat{\mu}_\tc{i}(t) = e^{\ii \Omega t} \hat{\sigma}_\tc{i}^++e^{-\ii \Omega t} \hat{\sigma}_\tc{i}^-$. {It is also possible to write this Hamiltonian explicitly in terms of the quantum controlled field sourced by each particle:
\begin{align}\label{eq:qcFieldH}
    H_\text{int}(t) =  \frac{\lambda}{2}\left(\hat{\mu}_\tc{a}(t) \hat{\phi}^{\text{qc}}_{\textsc{b}}(\mf z_\tc{a}(t))+\hat{\mu}_\tc{b}(t) \hat{\phi}^{\text{qc}}_{\textsc{a}}(\mf z_\tc{b}(t))\right),
\end{align}
where the field sourced by particle $\tc{I}$, $\hat{\phi}^{\text{qc}}_\tc{I}:\mathcal{M}\rightarrow \mathcal{H}_\tc{i}$ is defined as
\begin{align}\label{eq:qcField}
    \hat{\phi}^{\text{qc}}_\tc{i}(\mf x) &= \lambda \int \dd V' G_R(\mf x, \mf x')\hat{j}_\tc{i}(\mf x'),
\end{align}
and the operator valued operator current which sources the field is
\begin{align}
    \hat{j}_\tc{i}(\mf x') &= \chi_\tc{i}(t')\hat{\mu}_\tc{i}(t')\delta^{(3)}(\bm x' - \bm z_\tc{i}(t')).
\end{align}
In this sense, the Hamiltonian of Eq. \eqref{eq:qcFieldH} simply couples the current of one system along its trajectory with the field sourced by the other system. Notice that the field of Eq. \eqref{eq:qcField} does not posses any quantum degree of freedom, and is entirely determined by the quantum states of its sources. In other words, the Hilbert space in which $\hat{\phi}^{\text{qc}}_\tc{i}(\mf x)$ acts is not associated with the field at all---it is instead the Hilbert space of the degrees of freedom of the sources.}

The interaction picture unitary time evolution for the system is then given by
\begin{align}\label{eq:UIclass}
    \hat{U} &= \mathcal{T}\exp\left(-\ii \int \dd t \hat{H}_\text{int}(t)\right) \\&= \openone - \ii \int \dd t \hat{H}_\text{int}(t) + \mathcal{O}(\lambda^4).\nonumber
\end{align}
In the basis $\{\ket{g_\textsc{a}g_\textsc{b}},\ket{g_\textsc{a}e_\textsc{b}},\ket{e_\textsc{a}g_\textsc{b}},\ket{e_\textsc{a}e_\textsc{b}}\}$, such that $\hat{\sigma}_\textsc{i}^+ \ket{g_\textsc{i}} = \ket {e_\textsc{i}}$ and $\hat{\sigma}_\textsc{i}^+ \ket{e_\textsc{i}}= 0$, this unitary can be written to leading order in the coupling constant as
\begin{equation}\label{eq:classU}
    \hat{U} = \begin{pmatrix}
        0 & 0 & 0 & -\mathcal{M}_{\textsc{c}}^*\\
        0 & 0 & -\mathcal{N}^*_{\textsc{c}} & 0 \\
        0 & \mathcal{N}_{\textsc{c}} & 0 & 0\\
        \mathcal{M}_{\textsc{c}} & 0 & 0 & 0
    \end{pmatrix} + \mathcal{O}(\lambda^4),
\end{equation}
where
\begin{align}
    \mathcal{M}_\textsc{c} = -  \frac{\ii}{2} \lambda^2\int \dd t\dd t' e^{\ii \Omega(t+t')}\chi_\textsc{a}(t)\chi_\textsc{b}(t')\Delta(\mf z_\textsc{a}(t),\mf z_\textsc{b}(t')),\nonumber \\
    \mathcal{N}_\textsc{c} = -\frac{\ii}{2} \lambda^2\int \dd t\dd t' e^{\ii \Omega(t-t')}\chi_\textsc{a}(t)\chi_\textsc{b}(t')\Delta(\mf z_\textsc{a}(t),\mf z_\textsc{b}(t')). \label{eq:McNc}
\end{align}

If the two systems start in their ground state, \mbox{$\hat{\rho}_0 = \ket{g_\textsc{a}}\!\!\bra{g_\textsc{a}}\otimes\ket{g_\textsc{b}}\!\!\bra{g_\textsc{b}}$}, then, after the interaction, the state of the quantum systems is given by $\hat{\rho} = \hat{U}\hat{\rho}_0 \hat{U}^\dagger$. To fourth order in the coupling $\lambda$, we obtain
\begin{equation}\label{eq:rhoClass}
    \hat{\rho}_{\textsc{c}} = \begin{pmatrix}
        1 - |\mathcal{M}_{\textsc{c}}|^2 & 0 & 0 & \mathcal{M}_{\textsc{c}}^*\\
        0 & 0 & 0 & 0 \\
        0 & 0 & 0 & 0\\
        \mathcal{M}_{\textsc{c}} & 0 & 0 & |\mathcal{M}_{\textsc{c}}|^2
    \end{pmatrix} + \mathcal{O}(\lambda^6).
\end{equation}
We note that $\hat{\rho}_{\textsc{c}}$ above is a pure state due to the fact that the evolution of two quantum systems interacting through a {qc-field} is unitary. Hence, the final state of the systems can also be written as $\hat{\rho}_{\textsc{c}} = \ket{\psi}\!\!\bra{\psi}$, where the state vector $\ket{\psi}$ is given by
\begin{equation}
    \ket{\psi} = \frac{\ket{g_\tc{a}g_{\tc{b}}} + \mathcal{M}_{\textsc{c}}\ket{e_\tc{a}e_{\tc{b}}}}{\sqrt{1 + |\mathcal{M}_{\textsc{c}}|^2}} + \mathcal{O}(\lambda^6)
\end{equation}
to fourth order in $\lambda$.

One can also formulate a smeared version of the model of Eq. \eqref{eq:classHpointlike}. This is done by replacing the switching functions $\chi_\tc{i}(t)$ by spacetime smearing functions $\Lambda_\tc{i}(\mf x)$, which control both the space and time duration of the interaction, and also smear the Green's functions with respect to space. The resulting Hamiltonian for two smeared two-level quantum systems interacting via a relativistic quantum field can then be written as
\begin{align}\label{eq:classHsmearedfinal}
    &\hat{H}_\text{int}(t) = \frac{\lambda^2}{2}\int \dd t' \int \dd^3 \bm x_\textsc{a}\,\dd^3 \bm x_\textsc{b} \\
    &\Big(\hat{\mu}_\textsc{a}(t) \hat{\mu}_\textsc{b}(t'){\Lambda}_{\textsc{a}}(t,\bm x_\textsc{a}){\Lambda}_{\textsc{b}}(t',\bm x_\textsc{b})G_R(t,\bm x_\textsc{a};t',\bm x_\textsc{b})\nonumber\\
    &+\hat{\mu}_\textsc{b}(t)\hat{\mu}_\textsc{a}(t'){\Lambda}_{\textsc{b}}(t,\bm x_\textsc{b}) {\Lambda}_{\textsc{a}}(t,\bm x_\textsc{a})G_R(t,\bm x_\textsc{b};t',\bm x_\textsc{a}) \Big),\nonumber
\end{align}
where $\Lambda_\textsc{a}$ and $\Lambda_\textsc{b}$ denote the spacetime smearing function associated with the chosen energy eigenspace. 

The model of Eq. \eqref{eq:classHsmearedfinal} is the smeared version of the model of Subsection \ref{sub:classPointlike}, where the retarded Green's functions are smeared by the spacetime smearing functions $\Lambda_\tc{i}(\mf x)$ instead of simply the switching functions $\chi_\tc{i}(t)$. The unitary time evolution takes the shape of Eq. \eqref{eq:classU}, but in this case with the following expressions for $\mathcal{M}_{\textsc{c}}$ and $\mathcal{N}_{\textsc{c}}$,
\begin{align}
    \mathcal{M}_\textsc{c} = -  \frac{\ii}{2} \lambda^2\int \dd V \dd V'e^{\ii \Omega(t+t')}\Lambda_{\textsc{a}}(\mf x)\Lambda_{\textsc{b}}(\mf x')\Delta(\mf x,\mf x'),\nonumber  \\
    \mathcal{N}_\textsc{c} = -\frac{\ii}{2} \lambda^2\int \dd V\dd V' e^{\ii \Omega(t-t')}\Lambda_{\textsc{a}}(\mf x)\Lambda_{\textsc{b}}(\mf x')\Delta(\mf x,\mf x'),\label{eq:classMN}
\end{align}
which are now integrated over the spacetime volume corresponding to the four dimensional interaction regions. Notice that if $\Lambda_\tc{i}(\mf x) = \chi_\tc{i}(t)\delta^{(3)}(\bm x - \bm z_\tc{i}(t))$ we recover the exact expressions for $\mathcal{M}_\tc{c}$ and $\mathcal{N}_\tc{c}$ for the pointlike case. Under the assumptions that led to Eq. \eqref{eq:rhoClass}, the final state of the systems after the interaction is also given by Eq. \eqref{eq:rhoClass}, but with $\mathcal{M}_\tc{c}$ and $\mathcal{N}_\tc{c}$ given by Eq. \eqref{eq:classMN}.

\subsection{General localized quantum sources coupled to a relativistic field}\label{sub:classGeneral}

In this section we describe how general quantum sources can couple to a {qc-field} of arbitrary spin, using the theory developed in Section \ref{sec:classFields}. The sources will be described by localized quantum systems with a position degree of freedom, and possibly internal degrees of freedom in a finite dimensional Hilbert space, according to the general framework presented in~\cite{generalPD}. The position and momentum operators of the two respective systems will be denoted $\hat{\bm x}_\textsc{i}$ and $\hat{\bm p}_\textsc{i}$ for $\textsc{I} \in \{\textsc{A},\textsc{B}\}$. We will denote the algebraic basis of operators in the finite dimensional Hilbert space by $\{\hat{s}_{\textsc{i},i}\}$ where $i$ parametrizes a set of operators in $\mathbb{C}^n$. Thus, any observable in system $\tc{I}$ can be written as an algebraic combination of  $\hat{s}_{\textsc{i},i}$. Under these assumptions we can write the free Hamiltonians of each of the systems as $\hat{H}_{\textsc{i}}(\hat{\bm x}_\textsc{i}, \hat{\bm p}_\textsc{i}, \{\hat{s}_{\textsc{i},i}\})$.

In order to couple the systems with a field $\phi^a(\mf x)$, we will associate a current density $\hat{j}^{(\textsc{i})}_a(t,\hat{\bm x}_\textsc{i})$ to each system sourcing the field and acting on the $\mathbb{C}^n$ portion of the Hilbert space. In order to prescribe the interaction between the probes via the field, it is then enough to replace the classical currents in Eq. \eqref{eq:action} by their quantum counterparts, so that the interaction Hamiltonian in the position basis reads
\begin{align}\label{eq:classHgeneral}
    \hat{H}_\text{int}(t) = \frac{\lambda^2}{2}\int &\dd t' \int \dd^3 \bm x_\textsc{a}\,\dd^3 \bm x_\textsc{b} \ket{\bm x_\textsc{a}}\!\!\bra{\bm x_\textsc{a}}\otimes \ket{\bm x_\textsc{b}}\!\!\bra{\bm x_\textsc{b}}  \\
    &\Big(\hat{j}^{(\textsc{a})}_a(t,\bm x_\tc{a})\hat{j}^{(\textsc{b})}_b(t',\bm x_\textsc{b}) G^{ab}_R(t,\bm x_\textsc{a};t',\bm x_\textsc{b})\nonumber\\
    &+\hat{j}^{(\textsc{b})}_a(t,\bm x_\textsc{b})\hat{j}^{(\textsc{a})}_b(t,\bm x_\textsc{a}) G^{ab}_R(t,\bm x_\textsc{b};t',\bm x_\textsc{a}) \Big).\nonumber
\end{align}
This is the general formalism for two localized quantum systems interacting via a relativistic classical field.  Particular examples of the Hamiltonian of Eq. \eqref{eq:classHgeneral} can then be used to reproduce the interaction of atoms with electromagnetism (a classical version of the models of~\cite{Pozas2016,richard}), the interaction of nucleons with neutrinos (the classical version of the models of~\cite{neutrinos,carol}) and the coupling of quantum systems with classical gravity~\cite{remi,boris}. 

{{\subsection{Discussion: fields with no quantum degrees of freedom}\label{sub:classDiscussion}}

In this subsection we discuss the fields that participate in the quantum-controlled interaction. Recall that the field considered to mediate the interaction is devoid of any degrees of freedom, and is entirely sourced by the systems which couple to it. In fact, it was possible to prescribe the Hamiltonians of Eqs. \eqref{eq:classHpointlike}  and \eqref{eq:classHgeneral} entirely in terms of the propagator. On the other hand, it is possible to define fields that act on the Hilbert space of the sources. In the general case, the field sourced by a current $\hat{j}_a(\mf x)$ can be written as
\begin{equation}
    \hat{\phi}^a_{\tc{i}}(\mf x) = \lambda\int \dd t' \dd \bm x'_\tc{i} G_R^{ab}(\mf x, \mf x')\hat{j}^{(\tc{i})}_b(t',\bm x'_\tc{i}) \ket{\bm x'_\tc{i}}\!\!\bra{\bm x'_\tc{i}},
\end{equation}
which acts on the Hilbert space of each source. The quantum-controlled field could then be written as $\hat{\phi}_\text{qc}^a(\mf x) = \hat{\phi}^a_{\tc{a}}(\mf x)+\hat{\phi}^a_{\tc{b}}(\mf x)$. In this sense, the field would be a well-defined spacetime valued operator that acts on the Hilbert space of the sources. From the equation above, it is also clearly possible to see that the field does not possess its own degrees of freedom: it is entirely determined by the sources. Although it is devoid of degrees of freedom, the field \emph{does} obey equations of motion in the sense that
\begin{equation}
    \mathcal{P}[\hat{\phi}^\text{qc}_a(\mf x)] = \lambda\hat{j}^{(\tc{a})}_a(t,\bm x)\ket{\bm x}\!\!\bra{\bm x}_\tc{a}+\lambda\hat{j}^{(\tc{b})}_a(t,\bm x)\ket{\bm x}\!\!\bra{\bm x}_\tc{b},
\end{equation}
where $\mathcal{P}$ is the differential operator of Eq. \eqref{eq:P}. 



There are cases where the qc-field can be completely reduced to a classical field. Indeed, whenever either of the quantum systems which source the qc-field is in an eigenstate of the interaction Hamiltonian, it is possible to define a single classical field that takes a single value in each point of spacetime. However, in the the cases considered here, the Hamiltonian of the particles does not commute with the interaction Hamiltonian. Although some authors would refer to this as a `quantum superposition of fields', we will refrain from using this nomenclature, because there is no Hilbert space associated with the dynamics of the qc-field. 

Finally, we remark that the fields in this model are fully relativistic and causal. The compatibility with relativity is implemented by the retarded propagation of the field which is prescribed in the Hamiltonian of Eq.~\eqref{eq:classHgeneral}. The retarded propagation also prevents any faster-than-light signalling from one source to another, and makes this model suitable for analyzing quantum information protocols in relativistic setups.
}

\section{Quantum Systems Interacting via Quantum Fields}\label{sec:quant}

In this section we describe the interaction of localized quantum systems with a quantum field. That is, instead of prescribing a direct interaction which neglects the field's degrees of freedom, the systems will interact with the local degrees of freedom of a quantum field.

A general localized non-relativistic quantum system which is coupled to a quantum field is usually termed a particle detector~\cite{DeWitt,birrell_davies}, and the general framework for describing these systems has been widely studied in the literature, especially in the field of relativistic quantum information (RQI), where many protocols using particle detectors have been devised (e.g.,  entanglement harvesting~\cite{Valentini1991,Reznik1,reznik2}, entanglement farming~\cite{Farming}, quantum collect calling~\cite{Jonsson2,PRLHyugens2015} and quantum energy teleportation~\cite{teleportation,teleportation2014} to name a few).
In this section we will review the formulation of a general setup in which two detectors interact with a quantum field and compare it with the case where two localized quantum systems interact via a quantum-controlled classical field.

\subsection{The two-level UDW model coupled to a real scalar field}\label{sub:UDW}

The simplest particle detector model is the \mbox{two-level} Unruh-DeWitt (UDW) particle detector~\cite{Unruh1976,DeWitt}. It consists of a two-level quantum system which interacts locally with a scalar quantum field. Albeit simple, this model has been shown to reproduce some fundamental features of more complex and realistic systems, such as atoms interacting with the electromagnetic field~\cite{eduardoOld,richard} and gravitational~\cite{pitelli,boris} fields, as well as nucleons interacting with the neutrino fields~\cite{neutrinos,antiparticles}.

In more detail, the two-level UDW model consists of a detector, a field and an interaction between them. The quantum field is assumed to be a real scalar field which obeys an equation of motion of the form $\mathcal{P}[\phi(\mf x)] = 0$, where $\mathcal{P}$ is the linear differential operator that defines the field's equation of motion. Using the classical description for this free field, it is possible to build the local algebras of observables for the corresponding quantum field theory as follows. Consider the set of smooth compactly supported functions in Minkowski spacetime, $C_c^\infty(\mathcal{M})$, and the formal operators $\hat{\phi}(f)\in \mathcal{A}$ with $f\in C_c^\infty(\mathcal{M})$, where $\mathcal{A}$ is the $*$-algebra with an identity element generated by the $\hat{\phi}(f)$. We impose the commutation relations
\begin{equation}\label{eq:EISQUANTUM}
    \comm{\hat{\phi}(f)}{\hat{\phi}(g)} = \ii E(f,g),
\end{equation}
where $E(f,g)$ is the causal propagator distribution,
\begin{equation}\label{eq:E}
    E(f,g) = \int \dd V \dd V'E(\mf x, \mf x') f(\mf x) g(\mf x'),
\end{equation}
and where $E(\mf x,\mf x') = G_R(\mf x,\mf x') - G_A(\mf x,\mf x')$ is the retarded-\emph{minus}-advanced propagator associated with the differential operator $\mathcal{P}$. For example, one could obtain a massive Klein-Gordon field by considering $\mathcal{P} = \Box - m^2$, and its massless version by considering $\mathcal{P} = \Box$. For all purposes, one can then think of the quantum field $\hat{\phi}(f)$ as an operator-valued distribution, where the field $\hat{\phi}(\mf x)$ is formally defined by
\begin{equation}
    \hat{\phi}(f) = \int \dd V \,\hat{\phi}(\mf x) f(\mf x).
\end{equation}
For more details about this construction, we refer the reader to~\cite{Haag,kasiaFewsterIntro,ericksonCapacity} and references therein.

In the context of operator algebras, states are defined as positive functionals $\omega:\mathcal{A}\rightarrow \mathbb{C}$ which are normalized such that $\omega(\openone) = 1$. Then, the expected value of a field observable $\hat{A}\in\mathcal{A}$ is defined as $\langle{\hat{A}}\rangle_\omega \equiv \omega(\hat{A})$. A particularly useful class of states is that of Hadamard states, which are zero mean Gaussian states and such that the local behaviour of the field's two point function distribution $\omega(\hat{\phi}(\mf x) \hat{\phi}(\mf x'))$ has a particular singular structure~\cite{Haag,kayWald,fewsterNecessityHadamard}. Such states are also required in order to obtain a regular version of the stress-energy tensor, and can be seen as a generalization of a notion of vacuum states via the GNS construction (see e.g.~\cite{Haag,kayWald,fewsterNecessityHadamard,kasiaFewsterIntro}). The field's two-point Wightman function $W(\mf x,\mf x') = \langle{\hat{\phi}(\mf x) \hat{\phi}(\mf x')}\rangle_\omega$ can be decomposed as
\begin{equation}\label{eq:W}
    W(\mf x, \mf x')  = \frac{\ii}{2}E(\mf x,\mf x') + \frac{1}{2}H(\mf x,\mf x'),
\end{equation}
where $E(\mf x, \mf x')$ is the causal propagator and \mbox{$H(\mf x, \mf x') = \langle\{\hat{\phi}(\mf x),\hat{\phi}(\mf x')\}\rangle_\omega$} is the Hadamard distribution, which contains the state-dependent part of the Wightman function. We can also define the Feynman propagator $G_F(\mf x,\mf x') = \langle\mathcal{T}\hat{\phi}(\mf x)\hat{\phi}(\mf x')\rangle_\omega$, where $\mathcal{T}$ denotes the time ordering operation. Its real and imaginary parts can be written as
\begin{equation}\label{eq:GF}
    G_F(\mf x, \mf x') = \frac{\ii}{2}\Delta(\mf x,\mf x') + \frac{1}{2}H(\mf x,\mf x'),
\end{equation}
where $\Delta(\mf x, \mf x')$ is the retarded plus advanced propagator defined in Eq. \eqref{eq:deltaGRGA}  (which is independent of the state $\omega$). {One can then interpret $E(\mf x,\mf x')$ as the quantum state independent part of the propagators. It is fundamentally quantum since it comes from the commutation relations of Eq.~\eqref{eq:EISQUANTUM}. The state dependent distribution $H(\mf x,\mf x')$ represents the role played by the quantum state in the Wightman function and in the Feynman propagator. $\Delta(\mf x, \mf x')$ represents the classical symmetric propagation of the field between the events $\mf x$ and $\mf x'$. \tbb{Finally notice that it is possible to write the retarded and advanced propagators in therms of the Wightman function and the Feynman propagator as}}
\begin{align}
    \tbb{\ii G_R(\mf x, \mf x')} &\tbb{= W(\mf x, \mf x') - G_F^*(\mf x,\mf x'),}\\
    \tbb{\ii G_A(\mf x, \mf x')} &\tbb{= G_F(\mf x,\mf x') - W(\mf x, \mf x').}
\end{align}
\tbb{Notice that the real part of 
 $W(\mf f, \mf x')$ and $G_F(\mf x, \mf x')$ cancel, so that there is no state dependence in $G_R(\mf x, \mf x')$ and $G_A(\mf x, \mf x')$. For a summary of the distributions used throughout the manuscript, their description and the relationship between them, please refer to Table \ref{tab}.}

\begin{widetext}

\begin{table}[h!]
\centering
\color{black}
\begin{tabular}{ |Sc|Sc|Sc|Sc|Sc| } 

\hline
Distribution name & Symbol & In terms of $\hat{\phi}(\mf x)$ & Alternative Form & Description  \\
\hline
\hline
Retarded Green's function & $G_R(\mf x, \mf x')$ & Classical & $\ii (G_F^*(\mf x,\mf x') - W(\mf x , \mf x'))$ & Retarded propagation from $\mf x'$ to $\mf x$  \\
\hline
Advanced Green's function & $G_A(\mf x, \mf x')$ & Classical & $\ii (W(\mf x , \mf x') - G_F(\mf x, \mf x'))$ & Retarded propagation from $\mf x'$ to $\mf x$  \\
\hline
Symmetric propagator & $\Delta(\mf x, \mf x')$ & Classical & $G_R(\mf x, \mf x') + G_A(\mf x, \mf x')$ & Classical exchange between $\mf x$ and $\mf x'$  \\ 

\hline
\hline
Hadamard function & $H(\mf x, \mf x')$ & $\langle \{\hat{\phi}(\mf x),\hat{\phi}(\mf x')\}\rangle$ & $2\Re(W(\mf x,\mf x')),2 \Re(G_F(\mf x,\mf x'))$ & State dependent correlations  \\
\hline
Causal propagator & $E(\mf x, \mf x')$ & $-\ii \langle[\hat{\phi}(\mf x),\hat{\phi}(\mf x')]\rangle$ & $G_R(\mf x, \mf x') - G_A(\mf x, \mf x')$ & Commutator dependent correlations  \\ 

\hline
\hline
Wightman function & $W(\mf x, \mf x')$ & $\langle \hat{\phi}(\mf x)\hat{\phi}(\mf x')\rangle$ & $\tfrac{1}{2}H(\mf x, \mf x') + \tfrac{\ii}{2}E(\mf x, \mf x')$ & Field's correlation function  \\
\hline
Feynman propagator & $G_F(\mf x, \mf x')$ & $\langle \mathcal{T}\hat{\phi}(\mf x)\hat{\phi}(\mf x')\rangle$ & $\tfrac{1}{2}H(\mf x, \mf x') + \tfrac{\ii}{2}\Delta(\mf x, \mf x')$ & Time-ordered correlation function \\ 
\hline
 
\end{tabular}

\caption{Table with the different distributions used throughout the paper and their descriptions. The distributions with ``Classical'' as an entry on the third column of the table are the functions that appear in the interaction of classical systems---notice that it is in principle possible to artificially write them as a function of $\hat{\phi}(\mf x)$, but the state dependent part from different propagators cancel. Notice that there are different conventions for the Green's functions and that different authors might use different conventions for each of the propagators summarized above.\label{tab}}
\end{table}

\color{black}
\end{widetext}

The interaction of a UDW detector with the scalar quantum field~\cite{Unruh1976,DeWitt} is prescribed by the interaction Hamiltonian\footnote{For a covariant description see for instance~\cite{eduardo,us}.}
\begin{equation}\label{eq:pointlikeUDW}
    \hat{H}_\text{int}(t) = \lambda \int \dd^n \bm x \Lambda(\mf x) \hat{\mu}(t)\hat{\phi}(\mf x),
\end{equation}
where $\hat{\mu}(t) = e^{\ii \Omega t}\hat{\sigma}^+ + e^{-\ii \Omega t} \hat{\sigma}^-$ is the monopole moment operator in the interaction picture and $\Lambda(\mf x)$ is the spacetime smearing function supported locally around the trajectory $\mf z(t)$. $\Lambda(\mf x)$ implements a spatial profile for the detector as well as a finite time duration.

In order to draw a parallel between communication protocols using the UDW model above and the classical model of Eq. \eqref{eq:classHsmearedfinal}, we consider two UDW detectors undergoing comoving inertial trajectories in Minkowski spacetime. Each detector interacts with the same quantum field $\hat{\phi}(\mf x)$. To each of these detectors, we associate a spacetime smearing function $\Lambda_\textsc{i}(\mf x)$ and a monopole operator $\hat{\mu}_\textsc{i}(t)$. We further assume that the detectors are identical, so that they have the same energy gap.

This description for the interaction of the field and the probes involves three quantum systems with three different Hilbert spaces: The Hilbert space of each of the detectors (isomorphic to $\mathbb{C}^2$) and the quantum field's Fock space. The interaction of the detectors with the field is then prescribed as the sum of the individual interaction Hamiltonians:
\begin{equation}\label{eq:HtwoUDW}
    \hat{H}_\text{int}(t) = \!\lambda \!\left(\int \!\!\dd^n \bm x \Lambda_\textsc{a}(\mf x) \hat{\mu}_\textsc{a}(t)\hat{\phi}(\mf x)\!+\!\!\int \!\!\dd^n \bm x \Lambda_\textsc{b}(\mf x) \hat{\mu}_\textsc{b}(t)\hat{\phi}(\mf x)\right)\!.
\end{equation}
In this setup the probes do not interact directly with each other, unlike the case of the model of Section \ref{sec:class}. That is to say that the interaction Hamiltonian of each detector acts trivially on the Hilbert space of the other one. 

We can now proceed to compute the final state of the detectors after the interaction with the field. The unitary time evolution operator for the system will be given by
\begin{equation}\label{eq:UIquant}
    \hat{U} = \mathcal{T}\exp\left(-\ii \int \dd t \hat{H}_\text{int}(t)\right),
\end{equation}
where $\mathcal{T}\exp$ denotes the time ordering exponential. $\hat{U}$ then admits a Dyson expansion of the form
\begin{equation}
    \hat{U}  = \openone + \hat{U}^{(1)} + \hat{U}^{(2)} + \hat{U}^{(3)} + \hat{U}^{(4)} +\mathcal{O}(\lambda^5),
\end{equation}
with
\begin{align}
    \hat{U}^{(1)} &= - \ii \int \dd t_1 \hat{H}_\text{int}(t_1),\\
    \hat{U}^{(2)} &= - \int \dd t_1 \dd t_2 \hat{H}_\text{int}(t_1)\hat{H}_\text{int}(t_2)\theta(t_1-t_2),
\end{align}
where all integrals range from $-\infty$ to $\infty$. In general, for $n>2$ we have
\begin{align}\label{eq:Ugeneral}
    \hat{U}^{(n)} = &(-\ii)^n\int \dd t_1 \dots \dd t_n \\&\theta(t_1-t_2)\!\left(\prod_{k=2}^{n-1} \hat{H}_\text{int}(t_k) \theta(t_{k}- t_{k-1})\right)\!\theta(t_n - t_{n-1})\nonumber.
\end{align}
Usually the calculations with two UDW detectors are performed to second order in $\lambda$, with few exceptions which go to fourth order~{\cite{Pozas-Kerstjens:2015}}. We will display the results to fourth order in $\lambda$ so that we can draw a natural comparison with the classical model presented in Section \ref{sec:class}.

We assume that the system starts in the uncorrelated state $\hat{\rho}_0 = \ket{g_\textsc{a}}\!\!\bra{g_\textsc{a}}\otimes \ket{g_\textsc{b}}\!\!\bra{g_\textsc{b}}\otimes \ket{0}\!\!\bra{0}$, where $\ket{0}$ denotes the Minkowski vacuum. The final state of the detectors-field system will then be given by $\hat{\rho} = \hat{U} \hat{\rho}_0\hat{U}^\dagger$. This interaction results in corrections for the detectors' state which mix different products of field operators evaluated around the trajectories $\tc{A}$ and $\tc{B}$. Once the interaction is completed, one can compute the final state of the two detectors system by tracing out the field degrees of freedom. Tracing out the field replaces the products of field operators with its correlation functions, which are then smeared by the detectors spacetime smearing functions. We obtain the following final state of the detectors system
{\begin{align}\label{eq:rhoQuant4}
    {\hat{\rho}_\textsc{d} = \begin{pmatrix}
       1- \mathcal{Y}  & 0 & 0 & \mathcal{M}^* + \Upsilon\\
        0 & \mathcal{L}_{\textsc{bb}} + \Xi_{\textsc{ba}} & \mathcal{L}_{\textsc{ab}}^* + \Pi_{\textsc{ab}}^* & 0\\
        0 & \mathcal{L}_{\textsc{ab}} + \Pi_{\textsc{ab}} & \mathcal{L}_{\textsc{aa}} + \Xi_{\textsc{ab}} & 0\\
        \mathcal{M} + \Upsilon^* & 0 & 0 & |\mathcal{M}|^2 + \Theta
    \end{pmatrix},}
\end{align}
{where $\mathcal{Y}$ is given by}
\begin{equation}
    {\mathcal{Y} = \mathcal{L}_{\textsc{aa}} + \mathcal{L}_{\textsc{bb}} + |\mathcal{M}|^2 + \Xi_{\textsc{ab}} + \Xi_{\textsc{ba}} + \Theta,}
\end{equation}}
or, to leading order in $\lambda$,
\begin{align}\label{eq:rhoQuant2}
    \hat{\rho}_\textsc{d} = \begin{pmatrix}
       1- \mathcal{L}_\tc{aa} - \mathcal{L}_\tc{bb}  & 0 & 0 & \mathcal{M}^*\\
        0 & \mathcal{L}_{\textsc{bb}}  & \mathcal{L}_{\textsc{ab}}^* & 0\\
        0 & \mathcal{L}_{\textsc{ab}}  & \mathcal{L}_{\textsc{aa}}  & 0\\
        \mathcal{M} & 0 & 0 & 0
    \end{pmatrix}+\mathcal{O}(\lambda^4).
\end{align}
where we have
\begin{align}
    \mathcal{L}_{\tc{ij}} &= \lambda^2 \int \dd V \dd V' \Lambda_\tc{i}(\mf x) \Lambda_\tc{j}(\mf x') e^{- \ii \Omega(t-t')} W(\mf x,\mf x'),\label{eq:Lij}\\
    \mathcal{M} &= -\lambda^2 \int \dd V \dd V' \Lambda_\tc{a}(\mf x) \Lambda_\tc{b}(\mf x') e^{\ii \Omega(t+t')} G_F(\mf x,\mf x'),\label{eq:M}
\end{align}
and the remaining terms are of order $\mathcal{O}(\lambda^4)$. 

At this stage one can already see that the operator $\hat{\rho}_\textsc{d}$ in Eq. \eqref{eq:rhoQuant4} can be written as
\begin{equation}
    \hat{\rho}_\textsc{d} = \hat{\rho}_{\textsc{c}} + \hat{\rho}_{\textsc{q}},
\end{equation}
where $\hat{\rho}_{\textsc{c}}$ is the final state obtained in Eq. \eqref{eq:rhoClass} and $\hat{\rho}_{\textsc{q}}$ are the quantum corrections to the previous result, which are associated to the presence of quantum degrees of freedom in the field. This comparison will be made more precise in Subsection~\ref{sub:comparison}, where we will discuss how one recovers the results of the classical model in Subsection \ref{sub:classPointlike} from the interaction of Eq. \eqref{eq:HtwoUDW}. 

\tbb{It is also interesting to interpret the final state of the detectors in terms of the decoherence that they experience when interacting with the quantum field. Due to chosen interaction Hamiltonian being proportional to $e^{\ii \Omega t}\hat{\sigma}^+ 
 +e^{\ii \Omega t}\hat{\sigma}^- = \cos(\Omega t) \hat{\sigma}_x - \sin(\Omega t) \hat{\sigma}_y$, so that the state of the detectors experiences an amplitude damping type of decoherence~\cite{nielsen_chuang}. Had we chosen a different coupling proportional to the detectors free Hamiltonian, we would obtain a dephasing type of decoherence, as described in detail in~\cite{refereeAsked}. However, this type of coupling would not allow the detectors to become entangled in the setups we consider, as in this case, the interaction with the field would commute with their free Hamiltonians, and would not generate non-trivial dynamics. For this reason we will focus on the interaction of the form~\eqref{eq:HtwoUDW} in order to study quantum information protocols with this model.}

\tbb{Finally, it is worth  interpreting the different terms of Eq. \eqref{eq:rhoQuant4} in terms of field mode and detector excitation/deexcitation terms. With respect to a given choice of vacuum, it is possible to write the free quantum field as}
\begin{equation}
    \tbb{\hat{\phi}(\mf x) = \int \dd^n \bm k \left(u_{\bm k}(\mf x)  \hat{a}_{\bm k}+u_{\bm k}^*(\mf x)  \hat{a}_{\bm k}^\dagger\right),}
\end{equation}
\tbb{where $u_{\bm k}(\mf x)$ is a set of modes (solutions to the homogeneous equation of motion for the field $\phi(\mf x)$) and the $\hat{a}_{\bm k}$ and $\hat{a}_{\bm k}^\dagger$ are creation and annihilation operators, whose action can be interpreted as exciting or deexciting the mode labelled by $\bm k$. In this context, the interaction Hamiltonian contains terms of the form $\hat{\sigma}^\pm_{\tc{a}/\tc{b}} \hat{a}$ and $\hat{\sigma}^\pm_{\tc{a}/\tc{b}} \hat{a}^\dagger$, which implement detector and field modes excitations/deexcitations. In the specific scenario in which we computed $\hat{\rho}_\tc{d}$ in Eq. \eqref{eq:rhoQuant4}, both detectors start in their ground state, but they can still affect the field through counter-rotating terms ($\hat{\sigma}^-_{\tc{a}/\tc{b}} \hat{a}$ and $\hat{\sigma}^+_{\tc{a}/\tc{b}} \hat{a}^\dagger$) due to the time dependence introduced by the switching functions, which effectively encode the energy deposited into the system when  the detectors-field interactions are switched on and off. This picture may be useful to give some interpretation for the physical origin of the entanglement acquired by the detectors
.}

\subsection{General Particle Detectors}\label{sub:generalUDW}

The model presented in Subsection \ref{sub:UDW} is the simplest particle detector model, in the sense that the detectors are qubits which couple linearly to a real scalar field. Generalizations of this model have been explored in the literature in many different contexts, such as the study of the interaction of atoms with light~\cite{eduardoOld,Pozas2016,richard} and the study of the interaction of nucleons with neutrinos~\cite{neutrinos,antiparticles,carol}. A general particle detector model has been described in~\cite{generalPD}. Here we briefly review such model and present the formalism for the coupling of two detectors to the same quantum field theory.

Following the discussion of Subsection \ref{sub:UDW} and in line with the model of~\cite{generalPD}, a particle detector will be   a localized non-relativistic quantum system with a position degree of freedom and internal degrees of freedom described in $\mathbb{C}^n$. These can be described in a Hilbert space of the form $L^2(\mathbb{R}^3)\otimes\mathbb{C}^n$. In order to couple with a field of arbitrary spin, $\hat{\phi}^a(\mf x)$ (where $a$ is any collection of Lorentz indices), we must prescribe operator valued currents $\hat{j}_a^{(\tc{i})}(t,\hat{\bm x}_\tc{i})$ for each detector which act on the $\mathbb{C}^n$ portion of the Hilbert space. $\hat{\bm x}_\tc{i}$ here denotes the position operator for each system. The description of a quantum field of general spin can be found in~\cite{Khavkine2015}. Assuming the detectors to be inertial and comoving in Minkowski spacetime, the interaction Hamiltonian of two detectors with the quantum field can be written as
\begin{equation}
    \hat{H}_\text{int}(t) = \lambda \left(\hat{j}_a^{(\tc{a})}(t,\hat{\bm x}_\tc{a})\hat{\phi}^a(t,\hat{\bm x}_\tc{a}) + \hat{j}_a^{(\tc{b})}(t,\hat{\bm x}_\tc{b})\hat{\phi}^a(t,\hat{\bm x}_\tc{b})\right).
\end{equation}
This interaction Hamiltonian can be obtained from the one presented in \cite{generalPD} adding the assumption of flat spacetime and inertial comoving trajectories for each detector. 

In order to draw a more direct comparison with the classical Hamiltonian of Eq. \eqref{eq:classHgeneral}, it is possible to expand the interaction Hamiltonian above in the position basis of each detector, yielding:
\begin{align}\label{eq:HquantGeneralPosBasis}
    \hat{H}_\text{int}(t) = \lambda \Big(&\int \dd^3 \bm x_\tc{a}\ket{\bm x_\tc{a}}\!\!\bra{\bm x_{\tc{a}}}\hat{j}_a^{(\tc{a})}(t,{\bm x}_\tc{a})\hat{\phi}^a(t,{\bm x}_\tc{a}) \\&+ \int \dd^3 \bm x_\tc{b}\ket{\bm x_\tc{b}}\!\!\bra{\bm x_{\tc{b}}}\hat{j}_a^{(\tc{b})}(t,{\bm x}_\tc{b})\hat{\phi}^a(t,{\bm x}_\tc{b})\Big).\nonumber
\end{align}
In order to recover the scalar coupling model of Subsection \ref{sub:UDW}, it is enough to make the replacements $j_a^{(\tc{i})}(t,\bm x)\longmapsto \Lambda(\mf x)\hat{\mu}_\tc{i}(t)$, $\hat{\phi}^a(\mf x) \longmapsto \hat{\phi}(\mf x)$, and to expand the interaction Hamiltonian in the eigenbasis of the free Hamiltonian of each detector, restricting the accessible Hilbert space to two relevant subspaces (see Appendix \ref{sub:classSmeared}).

\subsection{Comparing quantum fields and qc-fields}\label{sub:comparison}

At this stage, it is possible to make a general comparison between the models presented in Section \ref{sec:class} and in this section. 
Let us compare Eqs. \eqref{eq:rhoClass} and \eqref{eq:rhoQuant2} for the final state of the quantum systems in the qc-field and fully quantum cases, respectively. To leading order, there are two differences: the $\mathcal{L}_\tc{ij}$ terms only appear for the fully quantum case, and the $\mathcal{M}$ terms are replaced by $\mathcal{M}_\tc{c}$ in the qc-field model. 


These differences can also be phrased in terms of the distributions $E(\mf x, \mf x')$, $H(\mf x, \mf x')$ and $\Delta(\mf x,\mf x')$ that compose the real and imaginary parts of the Wightman function and the Feynman propagator in Eqs. \eqref{eq:W} and \eqref{eq:GF}. Indeed, if one could set the quantum state dependent part of the propagators to zero, $H(\mf x, \mf x')\longmapsto 0$, only the imaginary part of the Feynman propagator survives in $\mathcal{M}$. This would make $\mathcal{M}\longmapsto \mathcal{M}_\tc{c}$. By the same token, if one could also set the quantum state independent part of the propagator to zero ($E(\mf x, \mf x') \longmapsto 0$), one would then get rid of the contributions from $W(\mf x, \mf x')$, giving $\mathcal{L}_\tc{ij}\longmapsto 0$. That is, if one could neglect the quantum contributions from $E(\mf x, \mf x')$ and $H(\mf x,\mf x')$ in the propagators, and only keep the classical symmetric propagation between the sources $\Delta(\mf x, \mf x')$, one would recover the results of the quantum-controlled classical model. Later in the manuscript we will be able to identify the physical regimes where the contributions of $E(\mf x, \mf x')$ and $H(\mf x, \mf x')$ can be neglected, and where the quantum-controlled field can be seen as an approximation of a fully featured quantum field.

Another important difference between the quantum-controlled and fully quantum cases is related to unitarity. In the qc-model the full Hilbert space is given by the Hilbert space of the quantum sources. This implies that the time evolution of the sources is unitary, and that no information about the sources is lost through the interaction. However, in the coupling with a fully featured quantum field, the sources will get entangled with the field itself. This implies that information about the sources can (and in general will) be lost after they interact, because the field now decoheres them. Concretely, after interacting with a quantum field, the probes end up in a mixed state, with purity equal to $\Tr(\hat{\rho}_\tc{d}^2)  = 1 - 2(\mathcal{L}_\tc{aa} + \mathcal{L}_\tc{bb})+\mathcal{O}(\lambda^4)$. Notice that the loss in purity of the detectors at leading order is given by the vacuum excitation probability of each detector, which is only present in the quantum field model. 



Remarkably, the models have one important feature in common: both are compatible with the notion of causality given by relativity. This implies that neither of the models allows for faster than light signalling and both respect the underlying spacetime causality\footnote{Although the smeared models contain causality violating issues even for relativistic fields, these issues are of the order of the detector size, and are well understood to stipulate a limit of validity of the non-relativistic quantum theory for the detectors~\cite{us2,PipoFTL}.}. However, while both models are relativistically causal and local in the spacetime sense, the qc-model implements a direct interaction between the two quantum systems. This is in contrast to the quantum field theoretic approach, where the probes interact through an intermediary system (the quantum field). \tbb{For a pictorial representation of the interaction Hamiltonians and regions of interaction for each of the models considered see Appendix B \ref{app:figures}.} 

Overall, the main distinctions between the qc-model and the quantum field model are associated with the intrinsically quantum properties of the field (e.g., spontaneous vacuum excitation, decoherence through the field, etc.). Although different, we will later see that there are regimes in which the quantum field can be well modelled by a qc-field. We will dedicate the remainder of this paper to 1) identify these regimes and 2) study how the differences between the models affect several relativistic quantum information protocols.


\section{Communication Protocols}\label{sec:communication}

Communication protocols that are mediated by quantum fields have been extensively studied in the literature (See, among others,~\cite{Landulfo,Barcellos,Casals,Jonsson1,Jonsson2,Jonsson3,PRLHyugens2015,Jonsson4,Simidzija_2020,ericksonCapacity,KojiCapacity}). In this section we briefly review some of the results in the literature when the protocol consists of two parties, and compare the results when the mediating field is quantum versus the case where it is a {qc-field}. We will focus on the case where the detectors are two level systems, as communication protocols become simpler, and often more intuitive with this choice.

\subsection{Quantum collect calling}

{In this section we will consider the communication protocol named ``quantum collect calling'' which was introduced in~\cite{Jonsson2}. This protocol involves the transmission of classical information between a sender (Alice) and a receiver (Bob) that operate quantum emitters coupled to a field. We will first discuss the model presented in~\cite{Jonsson2}, where a quantum field is considered for the protocol. Next we will adapt the protocol to the case where a qc-field mediates the interaction between sender and receiver. The comparison between these two cases will allow us to study the role of quantum degrees of freedom in quantum collect calling.}

\subsubsection{{Quantum collect calling via a quantum field.}}\label{sub:commQuant}

{Following~\cite{Jonsson2}, one can describe the protocol of quantum collect calling by modelling sender and emitter using} the UDW model presented in Subsection \ref{sub:UDW} with $\Lambda_\tc{a}(\mf x) = \chi_\tc{a}(t) \delta^{(3)}(\bm x - \bm x_\tc{a})$ and \mbox{$\Lambda_\tc{b}(\mf x) = \chi_\tc{b}(t) \delta^{(3)}(\bm x - \bm x_\tc{b})$}, which defines detectors $\tc{A}$ and $\tc{B}$ as localized at $\bm x_\tc{a}$ and $\bm x_\tc{b}$ with switching functions $\chi_\tc{a}(t)$ and $\chi_\tc{b}(t)$, respectively. In agreement with standard convention in quantum information literature, we will assume that Alice has control over the qubit labelled $\tc{A}$ and Bob has control over qubit $\tc{B}$. For convenience, we will assume that Alice has a qubit in a pure state $\ket{\psi_\tc{a}} = \alpha_\tc{a} \ket{g_\tc{a}} + \beta_\tc{a} \ket{e_\tc{a}}$ and Bob starts with a qubit in the pure state $\ket{\psi_\tc{b}} = \alpha_\tc{b} \ket{g_\tc{b}} + \beta_\tc{b} \ket{e_\tc{b}}$. Alice and Bob will then both couple to a massless real scalar field, according to the interaction Hamiltonian of Eq. \eqref{eq:HtwoUDW}, and their intention is to send classical information from Alice to Bob. 

Assuming that Alice couples to the field before Bob, one can then define a quantum channel between Alice's system and Bob's, $\mathcal{E}:\mathcal{D}_\tc{a} \longrightarrow \mathcal{D}_\tc{b}$, where $\mathcal{D}_\tc{i}$ denotes the set of density states of system $\tc{I}$. If \mbox{$\hat{\rho}_0 = \ket{\psi_\tc{a}}\!\!\bra{\psi_\tc{a}}\otimes \ket{\psi_\tc{b}}\!\!\bra{\psi_\tc{b}}\otimes \hat{\rho}_{\phi}$} denotes the state of the Alice-Bob-field system before their couplings with the field, then the channel $\mathcal{E}$ maps the state $\ket{\psi_\tc{a}}\!\!\bra{\psi_\tc{a}}$ into the state
\begin{equation}
    \mathcal{E}(\ket{\psi_\tc{a}}\!\!\bra{\psi_\tc{a}}) = \Tr_{\tc{a},\phi}(\hat{U}\hat{\rho}_0\hat{U}^\dagger),
\end{equation}
where $\hat{U}$ is the time evolution operator of Eq. \eqref{eq:UIquant} and $\Tr_{\tc{a},\phi}$ denotes the partial trace over Alice's and the field's degrees of freedom, which yields a state in Bob's space of states $\mathcal{D}_\tc{b}$.


In~\cite{Jonsson2} the authors find a lower bound for the classical channel capacity of a general setup where Alice and Bob possess qubits coupled to a quantum field. Their strategy was to consider a binary asymmetric channel, where Alice could send a `0' by not coupling to the field, or a `1' by coupling to it. This allows the channel capacity to be bound from below from  Alice's influence on Bob's excitation probability. 

In general, the probability that Bob finds his qubit in the excited state, $\ket{e_\tc{b}}$, after the interaction can be written as
\begin{equation}
    P_{e,\tc{b}} = |\beta_\tc{b}|^2 + \mathcal{R}_\tc{b} + \mathcal{S}_{\tc{ab}},
\end{equation}
where $\mathcal{R}_\tc{b}$ is the portion of the excitation probability which is independent of Alice's coupling, and $\mathcal{S}_\tc{ab}$ is dependent on Alice's interaction with the field. The contributions in $\mathcal{R}_\tc{b}$ are due to Bob's switching of the interaction and the local noise of the quantum field at his site. $\mathcal{S}_\tc{ab}$ is the contribution due to signalling from Alice's qubit. The term $|\beta_\tc{b}|^2$ is Bob's excitation probability previous to any coupling with the quantum field. Then, in~\cite{Jonsson2} it is shown that the lower bound for the channel capacity can be approximated to leading order in the coupling constant by
\begin{equation}\label{eq:capacity}
    C_\mathcal{E} \sim  \frac{2}{\ln 2} \left(\frac{\mathcal{S}_{\tc{ab}}}{4|\alpha_\tc{b}||\beta_{\tc{b}}|}\right)^2 + \mathcal{O}(\lambda^6),
\end{equation}
where $\mathcal{S}_\tc{ab}$ is given by
\begin{align}\label{eq:SABq}
    \mathcal{S}_{\tc{ab}} = -4 \lambda^2 \int \dd t_1 \dd t_2 \chi_\tc{a}(t_1) \chi_\tc{b}(t_2) \Re(\alpha_\tc{a}^*\beta_{\tc{a}}e^{\ii \Omega t_1})\nonumber\\
    \times\Im\left(\alpha_\tc{b}^*\beta_{\tc{b}}e^{\ii \Omega t_2}\right)G_R(t_1,{\bm x_\tc{a}};t_2,\bm x_\tc{b}).
\end{align}
Notice that the integral above only depends on the retarded Green's function of the field. For this reason, the lower bound of the channel capacity is independent of the field's quantum state. Our goal in the next subsection is to check whether the estimate above for the channel capacity changes when one considers the interaction with a {qc-field} instead, and to check whether quantum degrees of freedom play any role in the information protocol described here.

\subsubsection{{Quantum collect calling using a quantum-controlled field}}\label{sub:commClass}

It is possible to consider the same communication protocol discussed above when the detectors are instead coupled to a {qc-field}, according to the formalism described in Subsection \ref{sub:classPointlike}. In this context, there is no degree of freedom for the field, and the initial state of the Alice-Bob system is $\hat{\rho}_0 = \ket{\psi_\tc{a}}\!\!\bra{\psi_\tc{a}}\otimes \ket{\psi_\tc{b}}\!\!\bra{\psi_\tc{b}}$. The quantum channel is then
\begin{equation}
    \mathcal{E}(\ket{\psi_\tc{a}}\!\!\bra{\psi_\tc{a}}) = \Tr_A(\hat{U} \hat{\rho}_0 \hat{U}^\dagger),
\end{equation}
where $\hat{U}$ here denotes the time evolution operator of Eq. \eqref{eq:UIclass}. Same as {in the case of a fully featured quantum field}, we can find a lower bound for the channel capacity from the excitation probability of Bob's qubit. In the case of the {qc-field}, there is no vacuum noise, and the term $\mathcal{R}_\tc{b}$ vanishes, so that the excitation probability can be written as
\begin{equation}
    P_{e,\tc{b}} = |\beta_\tc{b}|^2 +\mathcal{S}^{\text{class}}_{\tc{ab}}.
\end{equation}
To leading order in the coupling constant one finds 
\begin{align}
    \mathcal{S}^{\text{class}}_{\tc{ab}} =  &-\:\alpha_\tc{a}^*\beta_{\tc{a}}\alpha_\tc{b}^*\beta_{\tc{b}} \mathcal{M}_{\tc{c}}-\alpha_\tc{a}\beta_{\tc{a}}^*\alpha_\tc{b}\beta_{\tc{b}}^* \mathcal{M}^*_{\tc{c}}\\&+\alpha_\tc{a}^*\beta_{\tc{a}}\alpha_\tc{b}\beta_{\tc{b}}^* \mathcal{N}_{\tc{c}}+\alpha_\tc{a}\beta_{\tc{a}}^*\alpha_\tc{b}^*\beta_{\tc{b}} \mathcal{N}_{\tc{c}}^*,\nonumber
\end{align}
where $\mathcal{M}_{\tc{c}}$ and $\mathcal{N}_{\tc{c}}$ are defined in Eq. \eqref{eq:McNc}. Alice's leading order contribution to Bob's excitation probability can then be recast as 
\begin{align}
    \mathcal{S}^{\text{class}}_{\tc{ab}} = -2\lambda^2 \int \dd t_1 \dd t_2 \chi_\tc{a}(t_1) \chi_\tc{b}(t_2) \Re(\alpha_\tc{a}^*\beta_{\tc{a}}e^{\ii \Omega t_1})\nonumber\\
    \times\,{\Im\left(\alpha_\tc{b}^*\beta_{\tc{b}}e^{\ii \Omega t_2}\right)\Delta(t_1,\bm x_\tc{a};t_2,\bm x_\tc{b}). }
\end{align}
{In essence, the difference between the qc-model and the fully quantum model is the replacement \mbox{$G_R \longmapsto \frac{1}{2}\Delta = \frac{1}{2}(G_R + G_A)$} in Eq. \eqref{eq:SABq}. In particular, if Alice interacts with the field before Bob, and no signal can be sent from Bob to Alice (so the $G_A$ contribution vanishes), the channel capacity of the qc-model is half as much as the one of the quantum case. On the other hand, in the limit where both interactions happen for times much longer than the space separation between the interactions (for instance, with $\chi_\tc{a}\rightarrow 1$ and $\chi_\tc{b}\rightarrow 1$), we obtain $\mathcal{S}_{\tc{ab}}^\text{class} \rightarrow  \mathcal{S}_{\tc{ab}}$, so that both models yield the same lower bound for the channel capacity.}

In order to understand why, when Alice interacts with the field before Bob, the channel capacity of the classical case is half as much as the quantum case, let us look at the retarded propagator. It can be expressed in terms of the Feynman propagator and the Wightman function as
{
\begin{align}
    \ii G_R(\mf x ,\mf x') &= W(\mf x, \mf x') - (G_F(\mf x, \mf x'))^*\nonumber\\
    &= \frac{\ii}{2} E(\mf x, \mf x') +  \frac{\ii}{2} \Delta(\mf x, \mf x').
\end{align}
}%
{That is, in Eq. \eqref{eq:SABq} the appearance of the retarded Green's function is due to a combination of the quantum ($E(\mf x, \mf x')$) and classical ($\Delta(\mf x, \mf x')$) state independent propagators. The (quantum) term $E(\mf x, \mf x')$ is responsible for increasing the transmission of classical information in this protocol.}

{Overall, we saw that in this specific protocol it is possible to use a field with quantum degrees of freedom to better transmit classical information. However, in the following subsection we will analyze a case where the \mbox{qc-field} is always better at transmitting classical information than a fully featured quantum field.}


\subsection{A non-perturbative approach to channel capacities}\label{sub:commNonpert}

 In contrast to the previous Subsection where the interaction of the qubits with the field was weak, in this Subsection we turn our attention to the case where Alice and Bob strongly couple to the field for just an instant. This kind of coupling is usually called a delta-coupling, and can be treated non-perturbatively. This model can be obtained from the models presented in Subsections \ref{sub:classPointlike} and \ref{sub:UDW} by considering the spacetime smearing functions to be given by
\begin{equation}\label{eq:deltaCoupling}
    \Lambda_\tc{i}(\mf x) = \eta_\tc{i}\delta(t - t_\tc{i}) f_\tc{i}(\bm x),
\end{equation}
where $\eta_\tc{i}$ defines an energy scale for the coupling and the $\delta(t-t_\tc{i})$ factor sharply localized the interaction of detector $\tc{I}$ in the space slice $t = t_\tc{i}$. Because we want to consider the situation where Alice sends a message to Bob, we will assume that Alice's interaction happens first.

In~\cite{ericksonCapacity}, the classical channel capacity of the channel established between Alice and Bob when they interact via a delta coupling with a quantum field is computed analytically. The authors show that if the initial state of the Alice-Bob system is separable, $\hat{\rho}_{\tc{ab},0} = \hat{\rho}_{\tc{a},0}\otimes \hat{\rho}_{\tc{b},0}$, Bob's final state after the interaction and after tracing over Alice's and the field's degrees of freedom is given by
\begin{align}\label{eq:rhoBquant}
    \hat{\rho}_\tc{b} = &\left(\frac{1}{2} + \frac{\nu_\tc{b}}{2}\cos\left(2E(\Lambda_\tc{a},\Lambda_{\tc{b}})\right)\right) \hat{\rho}_{\tc{b},0} \\&+ \left(\frac{1}{2} - \frac{\nu_\tc{b}}{2}\cos\left(2E(\Lambda_\tc{a},\Lambda_{\tc{b}})\right)\right)\hat{\mu}_\tc{b}(t_\tc{b}) \hat{\rho}_{\tc{b},0}\hat{\mu}_\tc{b}(t_\tc{b}) \nonumber\\
    &- \frac{\ii\nu_\tc{b}}{2} \sin(2E(\Lambda_\tc{a},\Lambda_{\tc{b}}))\theta_\tc{a}(t_\tc{a}) \comm{\hat{\mu}_\tc{b}(t_\tc{b})}{\hat{\rho}_{\tc{b},0}},\nonumber
\end{align}
where $E(f,g)$ denotes the integrated causal propagator (see Eq. \eqref{eq:E}),
\begin{equation}
    \nu_\tc{b} = e^{-2\mathcal{L}_{\tc{bb}}},
\end{equation}
and $\theta_\tc{a}(t_\tc{a}) = \Tr(\hat{\mu}_\tc{a}(t_\tc{a}) \hat{\rho}_\tc{a})$, where $t_{\tc{a}}$ is the time of the interaction of Alice's detector with the field.

In~\cite{ericksonCapacity}, it is then shown that the classical channel capacity of the quantum channel $\hat\rho_{\textsc{a},0}\mapsto\hat\rho_{\textsc{b}}$  is given by
\begin{equation}\label{eq:capacityDelta}
    C_{\mathcal{E}} = H\left(\frac{1}{2} + \frac{\nu_\tc{b}}{2}\left|\cos(2\lambda^2E(\Lambda_\tc{a},\Lambda_\tc{b}))\right|\right)-H\left(\frac{1}{2} + \frac{\nu_\tc{b}}{2}\right),
\end{equation}
 where $H(x) = -x\log(x) -(1-x)\log(1-x)$. Notice that because Alice's interaction happens before Bob's, one can replace the causal propagator $E(\Lambda_\tc{a},\Lambda_\tc{b})$ by the retarded propagator. That is, we end up with
\begin{equation}
    C_{\mathcal{E}} = H\left(\frac{1}{2} + \frac{\nu_\tc{b}}{2}\left|\cos(2\lambda^2G_R(\Lambda_\tc{a},\Lambda_\tc{b}))\right|\right)-H\left(\frac{1}{2} + \frac{\nu_\tc{b}}{2}\right).
\end{equation}
In particular, if Alice's region of interaction ($\text{supp}(\Lambda_\tc{a})$) is spacelike separated from Bob's ($\text{supp}(\Lambda_\tc{b})$), then the channel capacity is zero, as there is no signalling between sender and receiver.

For comparison, let us now consider the case where Alice and Bob are coupled via a {qc-field} according to the coupling prescribed in Subsection \ref{sub:classPointlike},  with the choice of spacetime smearing function of Eq. \eqref{eq:deltaCoupling}. We also assume Alice's interaction to happen before Bob's at \mbox{$t = t_\tc{a}<t_{\tc{b}}$} according to an inertial time $t$. In this case, it is possible to show that if the initial state of Alice's and Bob's system is separable, $\hat{\rho}_{\tc{ab},0} = \hat{\rho}_{\tc{a},0}\otimes \hat{\rho}_{\tc{b},0}$, then Bob's state after the interaction is given by
\begin{align}
    \hat{\rho}_\tc{b} = &\cos^2(\Delta(\Lambda_\tc{a},\Lambda_{\tc{b}})) \hat{\rho}_{\tc{b},0}\label{eq:rhoBclass} \\&+ \sin^2(\Delta(\Lambda_\tc{a},\Lambda_{\tc{b}})) \hat{\mu}_\tc{b}(t_\tc{b}) \hat{\rho}_{\tc{b},0}\hat{\mu}_\tc{b}(t_\tc{b}) \nonumber\\
    &- \ii\sin(\Delta(\Lambda_\tc{a},\Lambda_{\tc{b}}))\cos(\Delta(\Lambda_\tc{a},\Lambda_{\tc{b}}))\theta_\tc{a}(t_\tc{a}) \comm{\hat{\mu}_\tc{b}(t_\tc{b})}{\hat{\rho}_{\tc{b},0}},\nonumber
\end{align}
Comparing the result of Eq. \eqref{eq:rhoBquant} with Eq. \eqref{eq:rhoBclass}, we see\footnote{In order to compare Eq. \eqref{eq:rhoBquant} and Eq. \eqref{eq:rhoBclass}, notice that $\cos^2(\theta) = \tfrac{1}{2} + \tfrac{1}{2}\cos(2\theta)$, $\sin^2(\theta) =\tfrac{1}{2} - \tfrac{1}{2}\cos(2\theta)$, $\sin(2\theta) = 2 \sin(\theta) \cos(\theta)$ and that if Alice interacts with the field with $t_\tc{a}<t_\tc{b}$, then \mbox{$E(\Lambda_\tc{a},\Lambda_\tc{b})=G_R(\Lambda_\tc{a},\Lambda_\tc{b})=\Delta(\Lambda_\tc{a},\Lambda_\tc{b})$}.} that the channel capacity in the quantum-controlled case corresponds to the channel capacity in the quantum case setting $\nu_\tc{b} = 1$, or, equivalently, by setting the vacuum excitation probability to $\mathcal{L}_{\tc{bb}} = 0$. In particular, the channel capacity of this quantum channel is given by Eq. \eqref{eq:capacityDelta} evaluated at $\nu_\tc{b} = 1$:
\begin{equation}
    C_{\mathcal{E}}^{\text{class}} = H\left(\frac{1}{2} + \frac{1}{2}\left|\cos(2\lambda^2G_R(\Lambda_\tc{a},\Lambda_\tc{b}))\right|\right),
\end{equation}
where we used $H(1) = 0$. 

We can now compare the channel capacity {in the cases where a fully featured quantum field or a quantum-controlled classical field mediates the interaction}. It is possible to show that the channel capacity in Eq. \eqref{eq:capacityDelta} is monotonically increasing with $\nu_\tc{b}$, which always satisfies $\nu_\tc{b}\leq 1$. This implies that we always have $C_{\mathcal{E}}^{\text{class}} > C_\mathcal{E}$: delta coupled interactions mediated by {qc-fields} are always better at transmitting classical information than when mediated by quantum fields. This can be traced back to the fact that a true quantum field has its own quantum degrees of freedom, which become entangled with the probes and produce noise. This noise can harm communication protocols, which can be seen from the fact that the channel capacity in Eq. \eqref{eq:capacityDelta} is monotonically decreasing as a function of the vacuum excitation probability $\mathcal{L}_\tc{bb}$. 


While one cannot switch off the quantum degrees of freedom of a field that is fundamentally quantum, the analysis of this section shows that when the vacuum effects of the field are negligible, the effect of the quantum degrees of freedom of the field can also be neglected, at least for the purpose of classical communication.

\section{Entangling Protocols: Entanglement Acquired by two interacting quantum systems}\label{sec:entangling}

Two quantum systems that are initially uncorrelated can become entangled via an interaction. Most known interactions between quantum systems are fundamentally mediated by quantum fields. However, we do not always have to go down to quantum field theory to describe the interaction of two quantum systems. For example, to describe two spins interacting via the electromagnetic field one does not usually describes the field with quantum electrodynamics. Instead, it is common to replace the mediator by an effective direct interaction (e.g., spin-spin coupling), which is also able to entangle the systems. In this section we analyze the difference in the entanglement acquired by probes which are coupled to a quantum field or directly via the interaction described by a quantum-controlled classical field, according to the model of Section \ref{sec:class}. 

The study of entanglement acquired between two quantum systems coupled to a field in relativistic setups has been extensively studied in the literature, mostly in the context of quantum field theory~\cite{Valentini1991,Reznik1,reznik2,Salton:2014jaa,Pozas-Kerstjens:2015,HarvestingQueNemLouko,Pozas2016,HarvestingSuperposed,Henderson2019,bandlimitedHarv2020,ampEntBH2020,mutualInfoBH,threeHarvesting2022,twist2022}. Among the discoveries resulting from this line of studies is the fact that even causally disconnected probes can become entangled via the interaction of a quantum field, which gave rise to the entanglement harvesting protocol. In essence this protocol allows probes to extract entanglement previously existing in a quantum field. On the other hand, it is well known that a classical field cannot entangle two spacelike separated probes, because the field cannot contain previous quantum correlations. The main goal of this section is to precisely quantify the difference between {qc-fields} and quantum fields in entangling protocols.

\subsection{Entanglement via communication through a classical field}\label{sub:entClass}

In this section we consider two pointlike two-level quantum systems directly coupled according to the quantum-controlled classical model described in Subsection \ref{sub:classPointlike}. In this case we have seen that if the detectors start in the ground state, the final state of the system is given by Eq. \eqref{eq:rhoClass}. In particular, we see that the relevant matrix elements of the final state are proportional to the $\mathcal{M}_\tc{c}$ term, which is given by an integral of the propagator $\Delta(\mf x,\mf x')$. The first conclusion that can be drawn from this description is that if the interactions of detectors $\tc{A}$ and $\tc{B}$ are causally disconnected, then the detector's state is unaffected. This is because the retarded and advanced Green's functions $G_R(\mf x, \mf x')$ and $G_A(\mf x,\mf x')$ only have support  when the events $\mf x$ and $\mf x'$ are causally connected. In particular, for a massless field in a spacetime that respects the strong Huygens's principle~\cite{Hyugens1,RayHyugens,Huygens2}, $G_R(\mf x, \mf x')$ and $G_A(\mf x,\mf x')$ are only non-zero when $\mf x$ and $\mf x'$ are lightlike separated, and there is no ``leakage'' of the propagators inside the lightcone. In this case, two detectors can only affect each other when interacting via a {qc-field} if their interactions are at some point lightlike separated. Overall, we can establish an important result regarding the quantum-controlled classical model: detectors can only become entangled if their interactions with the field are causally connected.

It is possible to quantify the entanglement acquired by the detectors via communication through the propagation of the {qc-field}. In order to quantify the entanglement of the final state of Eq. \eqref{eq:rhoClass} it is convenient to choose the negativity as an entanglement quantifier. The negativity is a faithful entanglement quantifier for bipartite two-level systems, and can be used even for mixed states, which will make the comparison between {the quantum-controlled and truly quantum cases} simpler later on. The negativity of a bipartite state $\hat{\rho}$, $\mf{N}(\hat{\rho})$, is defined as the absolute value of the sum of the negative eigenvalues of the partial transpose (with respect to one of the partitions) of $\hat{\rho}$. The partial transpose of the state of Eq. \eqref{eq:rhoClass} has a single negative eigenvalue (if $\mathcal{M}_\tc{c}\neq 0$), so that its negativity reads, to leading order in $\lambda$,
\begin{equation}\label{eq:NegC}
    \mf{N}(\hat{\rho}_\tc{c}) = |\mathcal{M}_\tc{c}|.
\end{equation}
From the expression above we can also confirm that when the detectors' region of interaction are not causally connected, they will also not be entangled (see the definition of $\mathcal{M}_\tc{c}$ in Eq. \eqref{eq:McNc}).

\begin{figure}[h!]
    \centering
    \includegraphics[width=8.6cm]{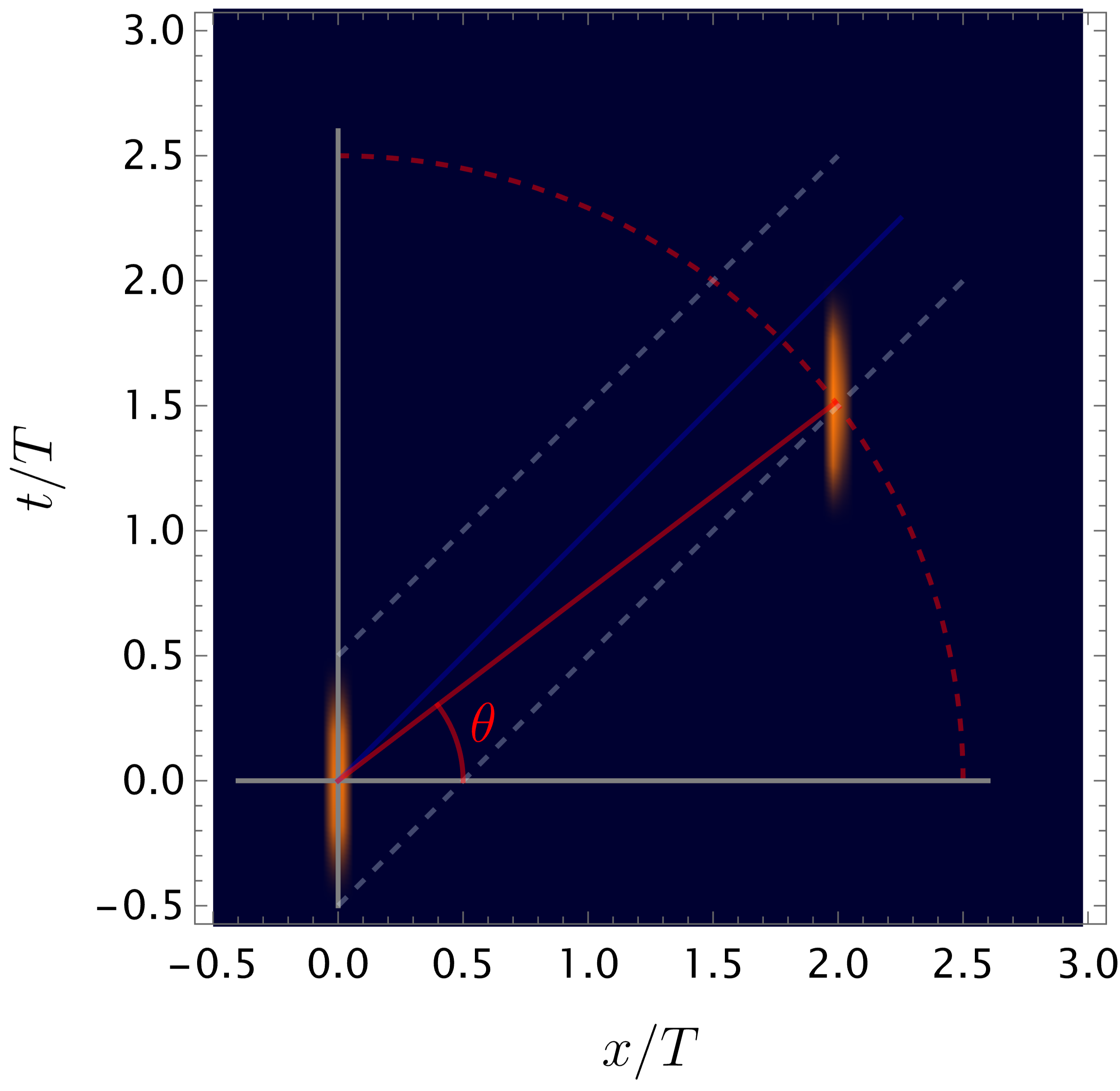}
    \caption{Setup for the configuration of the detectors as a function of the angle $\theta$.}
    \label{fig:SetupTheta}
\end{figure}

In order to consider a concrete example, let us consider a massless real scalar field in Minkowski spacetime and a specific spatial and temporal profile for the detectors, which undergo inertial comoving trajectories separated by a distance $L = |\bm L|$, where $\bm L$ is the separation vector between them. We will first prescribe the spacetime smearing functions as
\begin{align}
    \Lambda_\tc{a}(\mf x) &= \chi_T(t) \delta^{(3)}(\bm x),\label{eq:LambdaA}\\
    \Lambda_\tc{b}(\mf x) &= \chi_T(t-t_0) \delta^{(3)}(\bm x - \bm L),\label{eq:LambdaB}
\end{align}
where
\begin{equation}\label{eq:chiGaussian}
    \chi(t) = e^{-t^2/T^2}.
\end{equation}
With these choices the detectors are pointlike, $t_0$ is the time delay between the switchings, and $T$ controls the time duration of the interactions. The interaction of detector $\tc{A}$ is centered at the origin of the coordinate system, and the interaction of detector $\tc{B}$ is centered at the event $(t_0,\bm L)$. This choice makes the interaction not-compactly supported. The main consequence of this choice is that in principle the detectors will always be in causal contact due the tails of the Gaussians. However, 99.9999\% of the area of the detector's switching function is concentrated in an interval of width $7T$ centered at the Gaussian peak. We then define the interval $[t_m -3.5T,t_m+3.5T]$ as the strong support of the Gaussian, where $t_m$ is its peak value. As we will see, signalling outside that region will be negligible compared to the effect of the interaction when the strong supports are lightlike separated.

\begin{figure}[h]
    \centering
    \includegraphics[width=8.6cm]{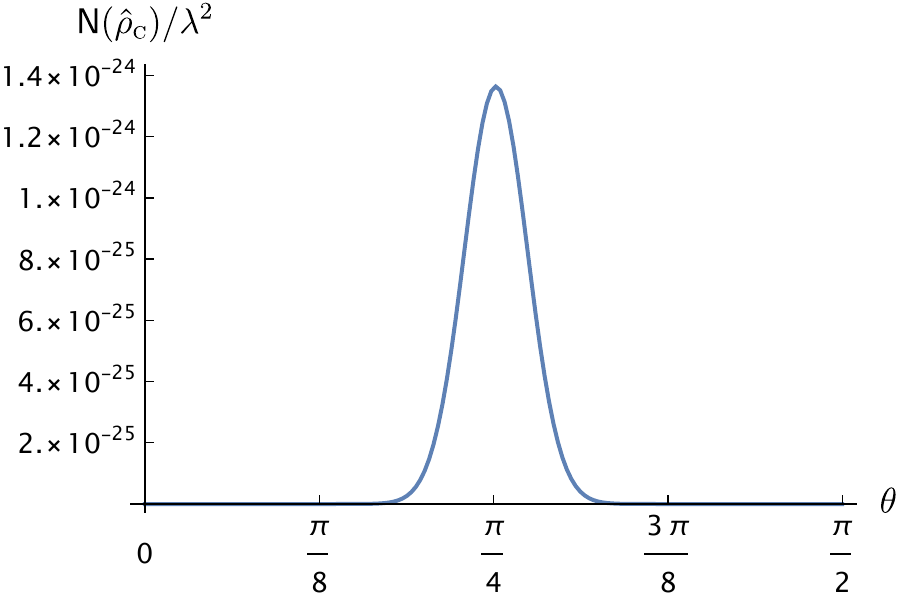}
    \caption{Plot of the negativity in the detectors state as a function of the angle $\theta$ for $\Omega T = 10$, $L = 10T$ for the Gaussian switching functions.}
    \label{fig:ClassicalNegativityGaussianTheta}
\end{figure}

We will analyze the negativity acquired by the detectors as we position detector $\tc{B}$ around different events of the form $(L\sin(\theta),L\cos(\theta),0,0)$ parametrized by the parameter $\theta\in(0,\pi/2)$, as shown in Fig. \ref{fig:SetupTheta}. In Fig. \ref{fig:ClassicalNegativityGaussianTheta} we plot the negativity in the detectors state as a function of $\theta$. We consider the distance between the detectors to be $L = 10T$, which ensures that the strong support of the Gaussians is spacelike separated. As we can see, there is no entanglement until a certain value of $\theta$, where the detectors stop being effectively spacelike separated. We then see a peak of the negativity when the \tbb{detectors' interaction regions} are lightlike separated at $\theta = \pi/4$. Notice that the plot is not completely symmetric with respect to the $\theta = \pi/4$ axis because the interaction regions are smeared in time, which slightly breaks the symmetry. 


\subsection{Entanglement through quantum fields}\label{sub:entQuant}

In this section we consider the entangling protocol outlined above in the case where the detectors are coupled to a real scalar \emph{quantum} field, following the interaction described in Subsection \ref{sub:UDW}. This protocol is usually termed entanglement harvesting, and has been thoroughly studied in the literature in many different scenarios~\cite{Valentini1991,Reznik1,reznik2,Salton:2014jaa,Pozas-Kerstjens:2015,HarvestingQueNemLouko,Pozas2016,HarvestingSuperposed,Henderson2019,bandlimitedHarv2020,ampEntBH2020,mutualInfoBH,threeHarvesting2022,twist2022}.  While usually the goal in entanglement harvesting is to extract entanglement from the field itself, here we will also be concerned with comparing our results with the ones obtained in the case where the detectors coupled to a {qc-field}. For this reason we will consider  regimes where the \tbb{detectors' interaction regions} are within causal contact and we will use the same choice of spacetime smearing functions for the probes as we did in Subsection \ref{sub:entClass}. Importantly, in the {fully featured quantum} case the detectors will be able to get entangled both via communication and by harvesting entanglement previously present in the background field.

Considering two UDW detectors initially in their ground states coupled to the vacuum state of a real scalar quantum field, the final state of the detectors to leading order will be given by Eq. \eqref{eq:rhoQuant2}. Same as in the previous section, we will use negativity (at leading order) to quantify the entanglement acquired by the detectors. For identical detectors interacting with the vacuum of Minkowski spacetime, the leading order negativity of the state $\hat{\rho}_\tc{d}$ is given by
\begin{equation}\label{eq:Nquant}
    \mf{N}(\hat{\rho}_\tc{d}) = \max(0,|\mathcal{M}| - \mathcal{L}),
\end{equation}
where $\mathcal{L} \equiv \mathcal{L}_{\tc{aa}} = \mathcal{L}_\tc{bb}$ for identical detectors. In this case the negativity reflects the competition between the non-local terms arising from $\mathcal{M}$ defined in Eq. \eqref{eq:M} and the local noise term $\mathcal{L}$ defined in Eq. \eqref{eq:Lij}. This local `vacuum' noise term is present only in the case where the field is quantum, as can be seen comparing Eqs.~\eqref{eq:NegC} and~\eqref{eq:Nquant}. Notice, however, that the negativity in~\eqref{eq:Nquant} can actually be larger than the negativity~\eqref{eq:NegC}, as we will see below. 

In order to draw a fair comparison between the {quantum-controlled and the truly quantum} models, we consider the same choice of spacetime smearing function of Eqs. \eqref{eq:LambdaA} and \eqref{eq:LambdaB}. Unlike the classical case, here we see that when the \tbb{detectors' interaction regions} are fully spacelike separated ($\theta\approx 0$) it is still possible for the detectors to become entangled. This is the entanglement that is extracted from the field, and not acquired by the detectors via communication. Namely, this is a truly quantum feature of the protocols of entanglement harvesting. We also see a peak when the interaction regions are lightlike separated. We do not see oscillations in the quantum field model because the real part of the propagator contributes out of phase with the imaginary part, yielding a non-zero value for the $\mathcal{M}$ term at all times.

\begin{figure}[h]
    \centering
    \includegraphics[width=8.6cm]{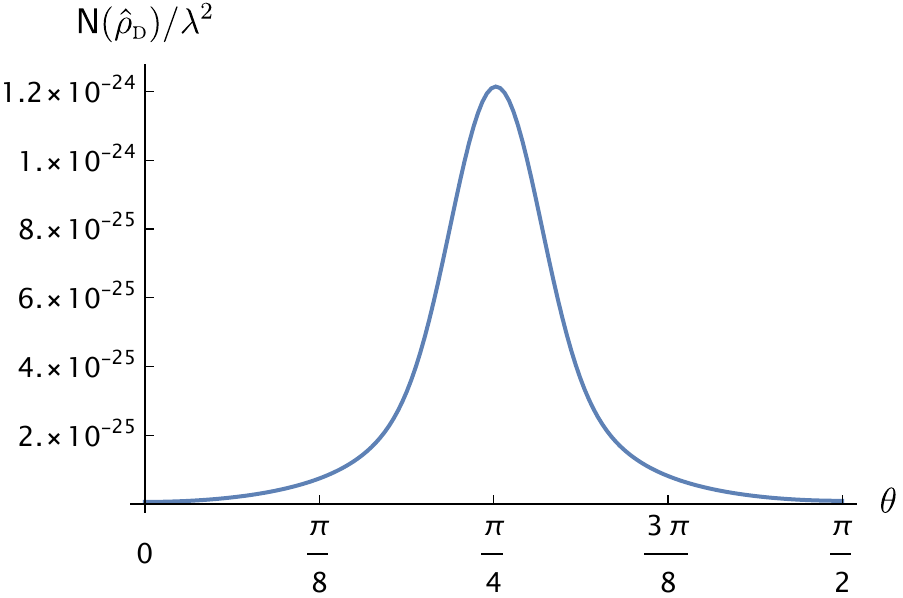}
    \caption{Plot of the negativity in the detectors state as a function of the angle $\theta$ for $\Omega T = 10$, $L = t_0 = 10T$ for Gaussian switching functions.}
    \label{fig:QuantumNegativityGaussianTheta}
\end{figure}

In Fig. \ref{fig:QuantumNegativityGaussianTheta} we fix $t_0 = 0$ (interactions happen simultaneously in the detector's frame) and show how the harvested entanglement varies with $\Omega$ for different values of $L$. The behaviour displayed in this plot is well known in the literature, where there is a threshold $\Omega$ so that the detectors become able to extract entanglement, after which the negativity peaks and slowly decreases with $\Omega$~\cite{Pozas-Kerstjens:2015,carol,hectorMass}. 

\begin{figure}[h]
    \centering
    \includegraphics[width=8.6cm]{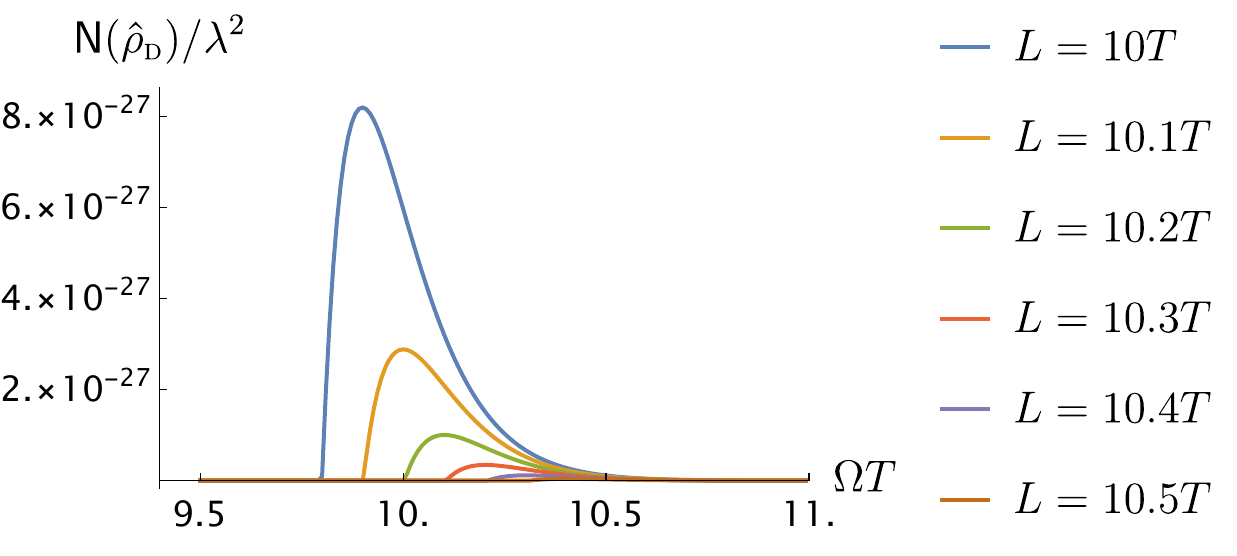}
    \caption{Plot of the negativity in the detectors state as a function of the detectors' gap for various values of the detector separation $L$ for Gaussian switching functions.}
    \label{fig:NvsOmegaGaussian}
\end{figure}

\tbb{We can also provide some physical intuition about the  origin of the entanglement acquired by the detectors. At leading order there is no real exchange of energy between detectors A and B through the field (this happens only at fourth order). In other words, at leading order there are no terms of the form $\hat{\sigma}_\tc{a}^-\hat{a}^\dagger$ times $\hat{\sigma}_\tc{b}^+\hat{a}$. That is, one needs a fourth order process in order for A and B to exchange energy through real emission-absorption processes. The leading order entanglement between the detectors is, therefore, acquired from the correlations of the field between the interaction regions, which can either be pre-existing (as in the case of spacelike entanglement harvesting) or be modified by the interaction of detector A (if A is in the causal past of B). In the latter case the entanglement is created through energy-less communication at leading order between A and B~\cite{Jonsson2,ericksonNew}. Something similar can also be seen in the quantum-controlled interaction, where the final state of the detectors is $\ket{\psi} \sim \ket{g_\tc{a}g_{\tc{b}}} + \mathcal{M}_\tc{c} \ket{e_\tc{a }e_\tc{b}}$ (Eq. \eqref{eq:rhoClass}), which is a superposition of the two detectors being in their ground and both detectors becoming excited, but does not contain any combinations that involve one detector being deexcited and another detector excited. In both models, energy-wise, the two detectors can become excited through the interaction with the field because the energy is coming by the time dependence of the interaction Hamiltonian: it is energy that is involved in switching the interaction on and off, as mentioned in Subsection~\ref{sub:UDW}.}

\section{When are the quantum degrees of freedom of the field negligible?}\label{sec:entComparison}

One legitimate question that can be asked about the {quantum-controlled classical} model with quantum sources is whether it can reproduce the phenomenology of the fully quantum model in some regimes. If the {qc-field} model is to hold any physical value it should indeed be able to reproduce the same physics as the fully quantum model in the regimes where the quantum features of the field do not play any relevant role. To answer this question, in this section we will compare the two models studied in Subsections \ref{sub:entClass} and \ref{sub:entQuant}, giving special attention to the regimes where the quantum field case can be well approximated by the {quantum-controlled} case. 

We choose to do this comparison for the the study of the entanglement acquired by two detectors when they interact with the field. The question of whether a model where the field is not fully quantum can predict that two systems that interact with the field get entangled is certainly relevant~\cite{MVPRD2020}, and as we will see this comparison already showcases the differences that will appear in any other more general protocol when considering two quantum systems communicating through a quantum field.

The scales relevant for addressing the regimes where qc-fields can approximate quantum fields are the detectors spatial separation $L$, their time separation, $t_0$, their energy gap $\Omega$ and the time of their switching, $T$. The relevant dimensionless parameters are then $L/T$, $t_0/T$ and $\Omega T$. It is already known that as $L/T$ increases past $t_0/T$ and as $t_0/T$ increases past $L/T$, the entanglement acquired by the detectors decreases (the further from light contact, the less entanglement between the detectors there will be in both models), so that the optimal rate $L/t_0$ is approximately $1$ making the detectors approximately lightlike separated. We also saw that the quantum field case can feature entanglement even when the \tbb{detectors' interaction regions} are spacelike separated, which is impossible in the {quantum-controlled} case. {In this sense, one of the conditions that is required for the quantum case to reduce to {the model with no quantum degrees of freedom} is that the detectors have to be causally connected}. This imposes a restriction on the parameters $L/T$ and $t_0/T$, so that the quantum field case can be well approximated by the {quantum-controlled classical} case. The study that remains to be conducted is what are the conditions over $\Omega T$ which allow the {fully featured} quantum field case to be well modelled by the {quantum-controlled} field scenario.

As mentioned in Subsection \ref{sub:comparison}, a main difference between {the cases where the field has quantum degrees of freedom or not} is the fact that {fully featured} quantum fields can produce local noise excitations in the detectors. This local noise is a consequence of the detectors becoming entangled with the field itself, which decoheres the state of the detectors. The decoherence results in a decrease of the entanglement between the detectors, as they share part of the entanglement with the field. This can be seen in Eq. \eqref{eq:Nquant} for the negativity of the detectors, where we see that the vacuum noise $\mathcal{L}$ contributes negatively to the entanglement acquired by them. It is then clear that a condition so that the {true} quantum case can be mimicked by the {quantum-controlled case} is that the $\mathcal{L}$ term is much smaller than the nonlocal $\mathcal{M}$ term. This condition can be achieved in the case where $\Omega T\gg 1$, or, in other words, in the limit where the interaction time is much larger than the characteristic time scale of the detectors.


\begin{figure}[h!]
    \centering
    \includegraphics[width=8.9cm]{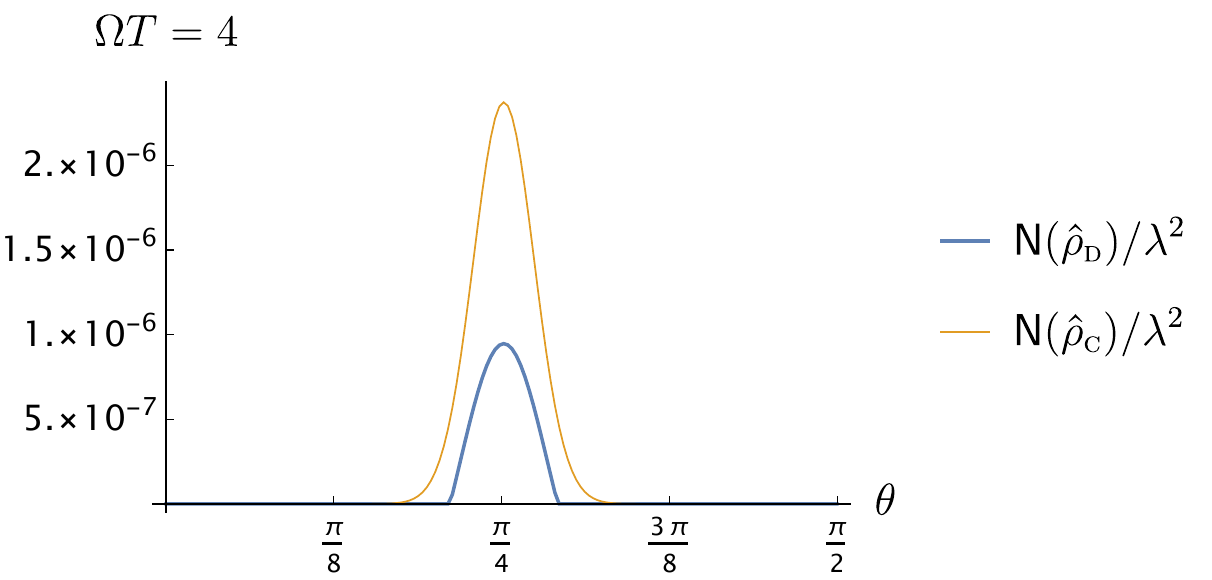}
    \includegraphics[width=8.9cm]{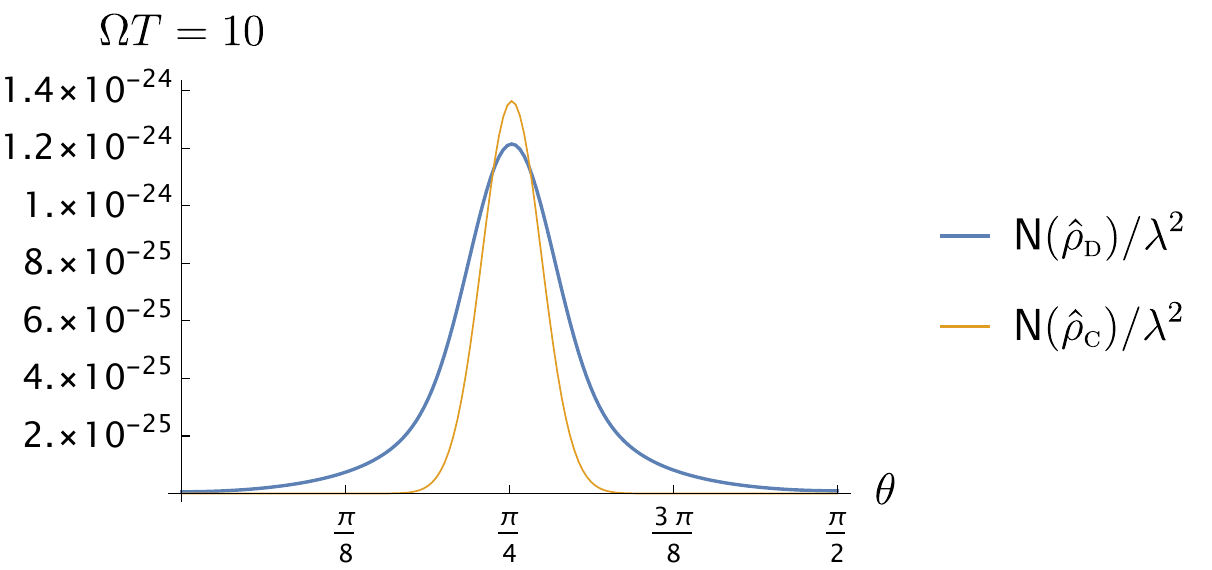}
    \includegraphics[width=8.9cm]{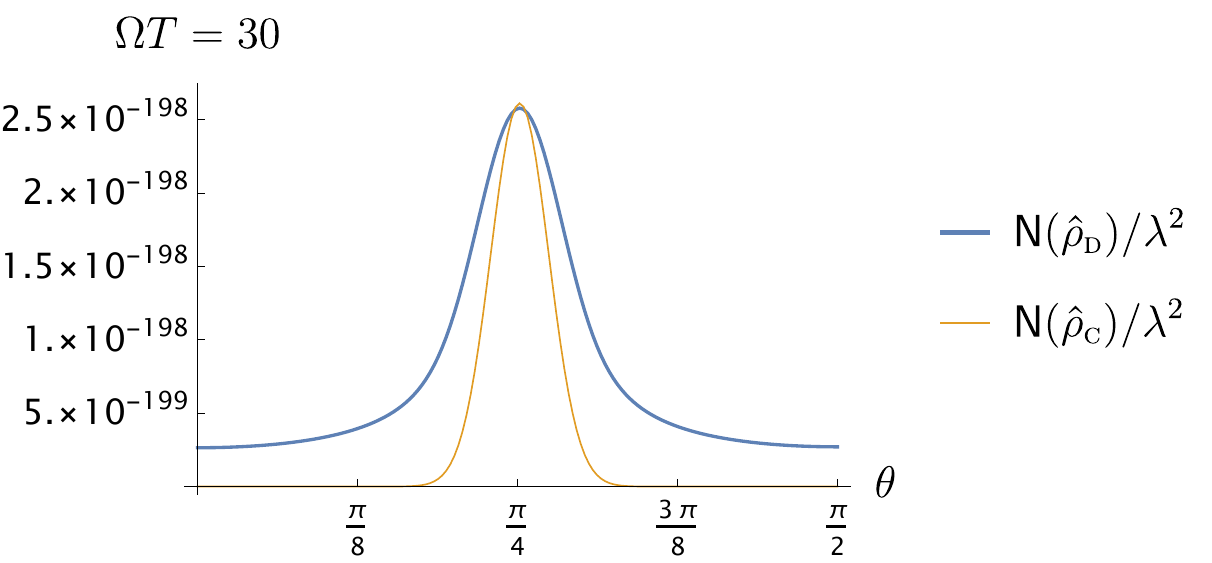}
    \includegraphics[width=8.9cm]{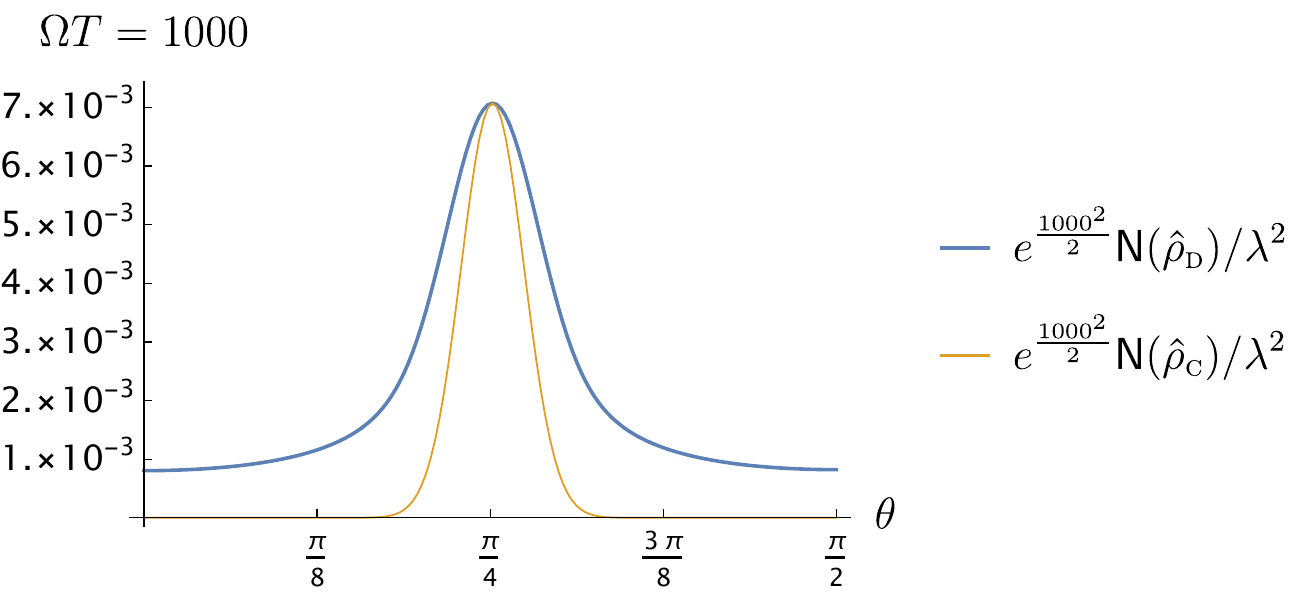}
    \caption{Plot of the negativity in the detectors state for both the classical and quantum cases as a function of the angle $\theta$ for $L = t_0 = 10T$ and multiple values of $\Omega T$ for Gaussian supported switching functions.}
    \label{fig:VaryingOmegaClassicalQuantumGaussian}
\end{figure}

Another condition for the quantum field and {qc-field} models to behave similarly is that $\mathcal{M} \approx \mathcal{M}_\tc{c}$. That is, that the imaginary part of the propagator contributes significantly more to the $\mathcal{M}$ term than its real part. In Fig. \ref{fig:VaryingOmegaClassicalQuantumGaussian} we show plots for the negativity as a function of $\theta$ for the setup of Fig. \ref{fig:SetupTheta} for different values of $\Omega T$ considering both the Gaussian switching of Eq. \eqref{eq:chiGaussian}. We see that when the \tbb{detectors' interaction regions} are in light contact it is possible to get more entanglement between the detectors when their interaction is via the {qc-field} than when the detectors interact via the quantum field. As we mentioned earlier, this is due to the local noise, which decreases the entanglement acquired by the detectors when they interact with a quantum field. That is, although we always have $|\mathcal{M}| \geq |\mathcal{M}_\tc{c}|$, we also have $\mathcal{L}>0$, which allows the negativity in the {quantum-controlled} case to surpass that of the quantum case when $\mathcal{L}$ is comparable to $|\mathcal{M}|$. In Fig. \ref{fig:VaryingOmegaClassicalQuantumGaussian} we also see that under the assumption that $\Omega T\gg 1$, the {quantum field and qc-field} models give similar predictions when the \tbb{detectors' interaction regions} are causally connected ($\theta \approx \pi/4$). 

Overall, we can conclude that the {fully} quantum case can be well modelled by the {quantum-controlled} case only if three conditions are satisfied: 1) the systems involved in the protocol must be causally connected, 2) the interaction time with the mediating field has to be much larger than the characteristic time scale of the detectors ($T \gg 1/\Omega$), and 3) the interactions with the field have to be sufficiently weak. {Condition 3 is necessary to avoid the major discrepancies between classical and quantum physics that take place for high energies, {which are not implemented simply by the retarded Green's function}.} 

Finally, notice that if the three conditions above are satisfied, then the density operator of Eq. \eqref{eq:rhoQuant2} obtained in the fully quantum case reduces to the density operator obtained in the {qc-model} in Eq. \eqref{eq:rhoClass}. Indeed, assuming the three conditions, we have $|\mathcal{L}_{\tc{ij}}|\leq \mathcal{L}\ll |\mathcal{M}|$. Then, to leading order in $\lambda$ we have
\begin{align}
    \hat{\rho}_\tc{d} = |\mathcal{M}|&\!\begin{pmatrix}
    (1 - 2{\mathcal{L}})/{|\mathcal{M}|} & 0 & 0& \!\!{\mathcal{M}^*}/{|\mathcal{M}|}\\
    0 & \!\!{\mathcal{L}}/{|\mathcal{M}|} & {\mathcal{L}_{\tc{ab}}}/{|\mathcal{M}|} & 0 \\ 
    0 & \!\!{\mathcal{L}_\tc{ba}}/{|\mathcal{M}|} & {\mathcal{L}}/{|\mathcal{M}|} & 0\\
    {\mathcal{M}}/{|\mathcal{M}|} & 0& 0& 0\end{pmatrix}\nonumber\\
    &\approx \begin{pmatrix}
    1  & 0 & 0 & {\mathcal{M}^*}\\
    0 & 0 & 0 & 0 \\ 
    0 & 0 & 0 & 0\\
    {\mathcal{M}} & 0& 0& 0\end{pmatrix},
\end{align}
which is the leading order result from Eq. \eqref{eq:rhoClass}. That is, the three assumptions discussed above ensure that the classical model can be used to approximate the interaction with the quantum field.

\section{Conclusions}\label{sec:conclusions}

We studied the difference between two models for fields which mediate interactions between quantum systems. Namely, we analyzed the differences between a fully featured quantum field theory and a theory where the field is stripped from all of its quantum features.

In more detail, we studied quantum information protocols which are mediated by {relativistic fields with or without quantum degrees of freedom}. We discussed the regimes in which a {quantum-controlled field (qc-field) theory (where the sources are quantum but the field has no quantum degrees of freedom)} can be used to model a process that is mediated by a quantum field. We also classified the regimes which display behaviour that explicitly depends on genuinely quantum features of the field. We then compared communication protocols and entangling protocols using the two different models for the field.

In the case of communication protocols, we found that for short time interactions, fields {without quantum degrees of freedom} would often outperform fully featured quantum fields at transmitting classical information. This is because,  for protocols that do not explicitly make use of the quantum nature of the field, the field itself gets entangled with the sender and receiver, producing an effective noise in communication channels between two systems. In our analysis, we explicitly showed that the channel capacity associated with {quantum-controlled} fields is indeed larger than the one associated with true quantum fields in specific scenarios. Nevertheless, for  protocols where the quantum features of the field play a relevant role, such as quantum collect calling, we also showed that quantum fields can yield a larger channel capacity.

We also studied entangling protocols mediated by {qc-fields} and quantum fields. We found that there are regimes in which a field with no quantum degrees of freedom can entangle two probes more than a quantum field can. This is due to the fact that when interacting with a quantum field, the probes also become entangled with the field itself, which decreases the entanglement between each other. We also verified the well known result that a quantum field can entangle spacelike separated probes~\cite{ericksonNew}, while a field with no quantum degrees of freedom cannot. This result is key to numerous protocols in relativistic quantum information such as entanglement harvesting and quantum energy teleportation~\cite{teleportation,teleportation2014}. 

Finally, we identified three conditions which are necessary for a relativistic quantum information protocol to {be independent of the quantum degrees of freedom of the field}. It is necessary that 1) the probes are causally connected, 2) that they interact for a time long enough 3) and that the interaction with the field is weak. These conditions set the regimes where a quantum field can be well modelled by a  qc-field with no quantum degrees of freedom. Our results could also  be relevant if one wishes to experimentally verify whether a field is fundamentally quantum or not. In this case, at least one of the conditions above must not be satisfied in an experimental setup, so that the experiment explores regimes where the qc-field model and the fully quantum one make different predictions.  

\begin{acknowledgements}\label{sec:ack}
T. R. P. acknowledges support from the Natural Sciences and Engineering Research Council
of Canada (NSERC) via the Vanier Canada Graduate Scholarship. E. M-M. is funded by the NSERC Discovery program as well as his Ontario Early Researcher Award. Research at Perimeter Institute is supported in part by the Government of Canada through the Department of Innovation, Science and Industry Canada and by the Province of Ontario through the Ministry of Colleges and Universities. Perimeter Institute and the University of Waterloo are situated on the Haldimand Tract, land that was promised to the Haudenosaunee of the Six Nations of the Grand River, and is within the territory of the Neutral, Anishnawbe, and Haudenosaunee peoples.
\end{acknowledgements}

\appendix

\twocolumngrid

\section{Smeared two-level systems coupled to a real scalar field}\label{sub:classSmeared}

{In this appendix, we will show one way in which one can obtain the smeared model of Eq. \eqref{eq:classHsmearedfinal} by considering two quantum sources with a position degree of freedom. Assume that quantum systems $\tc{A}$  and $\tc{B}$ are described each in a Hilbert space $\mathcal{H}\cong L^2(\mathbb{R}^3)\otimes \mathbb{C}^2$, where $\mathbb{C}^2$ is associated to their monopole and $L^2(\mathbb{R}^3)$ is associated to their position degree of freedom. Assume the free Hamiltonian of the systems to be given by
\begin{equation}
    \hat{H}_\textsc{i} = H^{\bm x}_\textsc{i}(\hat{\bm x}_\textsc{i},\hat{\bm p}_\textsc{i}) + \Omega\hat{\sigma}^+_\textsc{i}\hat{\sigma}^-_\textsc{i}
\end{equation}
for $\tc{I} = \tc{A}, \tc{B}$, where $H^{\bm x}_\textsc{i}(\hat{\bm x}_\textsc{i},\hat{\bm p}_\textsc{i})$ is a function of each system's position and momentum operators $\hat{\bm x}_\textsc{i}$ and $\hat{\bm p}_\textsc{i}$, and only acts on the $L^2(\mathbb{R}^3)$ portion of their respective Hilbert spaces. We further assume that the eigenfunctions of the Hamiltonians $H^{\bm x}_\textsc{i}$ are localized around the trajectories $\mf z_\textsc{i}(t)$, and have discrete energy levels (e.g. an atom or a harmonic oscillator where the potential is centered at $\mf z_\tc{i}(t)$).

In order to write the interaction of a quantum system with a classical field, we replace the classical current density \mbox{${j}^{(\textsc{i})}(\mf x)$} by a quantum current density \mbox{$\hat{j}^{(\textsc{i})}(t,\hat{\bm x}_\textsc{i}) = \chi_\textsc{i}(t)\hat{\mu}_\textsc{i}(t)f_\textsc{i}(\hat{\bm x}_\textsc{i})$}, which acts both in the particles' position degree of freedom via its functional dependence on the position operator $\hat{\bm x}_\textsc{i}$ and in the $\mathbb{C}^2$ portion of the Hilbert space via $\hat{\mu}_\textsc{i}(t)$. As in the previous example, we introduce switching functions $\chi_\textsc{i}(t)$. The interaction Hamiltonian can then be written in the interaction picture in the position basis as
\begin{align}
    &\hat{H}_\text{int}(t) = \frac{\lambda^2}{2}\int \dd t' \int \dd^3 \bm x_\textsc{a}\,\dd^3 \bm x_\textsc{b} \ket{\bm x_\textsc{a}}\!\!\bra{\bm x_\textsc{a}}\otimes \ket{\bm x_\textsc{b}}\!\!\bra{\bm x_\textsc{b}} \nonumber \\
    &\Big(\hat{\mu}_\textsc{a}(t)\hat{\mu}_\textsc{b}(t')\chi_\textsc{a}(t)\chi_\textsc{b}(t')f_\textsc{a}(
    \bm x_\textsc{a})  f_\textsc{b}(\bm x_\textsc{b}) G_R(t,\bm x_\textsc{a};t',\bm x_\textsc{b})\nonumber\\
    &\!+\hat{\mu}_{\textsc{b}}(t) \hat{\mu}_{\textsc{a}}(t') \chi_\textsc{b}(t) \chi_\textsc{a}(t')f_\textsc{b}(\bm x_\textsc{b}) f_\textsc{a}(\bm x_\textsc{a}) G_R(t,\bm x_\textsc{b};t',\bm x_\textsc{a}) \Big).
\end{align}
One can then expand these in terms of the energy levels of the Hamiltonians $\hat{H}_\textsc{i}^{\bm x}$. Let $\hat{H}_\textsc{i}^{\bm x} |{n^{(\textsc{i})}}\rangle = \mathcal{E}^{(\textsc{i})}_{n}|{n^{(\textsc{i})}}\rangle$ with $\psi_n^{(\tc{i})}(\bm x) = \langle{\bm x_\tc{i}}|{n^{(\tc{i})}}\rangle$ for $\textsc{I}  \in \{\textsc{A},\textsc{B}\}$, so that the interaction Hamiltonian reads
\begin{align}\label{eq:fullClassH}
    \hat{H}_\text{int}(t) \!= &\frac{\lambda^2}{2}\!\!\!\sum_{\substack{n,m\\n',m'}}\!\int \!\dd t' \!\!\int \!\dd^3 \bm x_\textsc{a}\dd^3 \bm x_\textsc{b} |{n^{(\tc{a})}}\rangle\!\langle{m^{(\tc{a})}}\!|\!\otimes\! |{n'{}^{(\tc{b})}}\rangle\!\langle{m'{}^{(\tc{b})}}\!|  \nonumber\\
    \times&\Big(\hat{\mu}_\textsc{a}(t) \hat{\mu}_\textsc{b}(t')e^{\ii(E^{(\textsc{a})}_{nm}t+E^{(\textsc{b})}_{n'm'}t')}\nonumber\\ 
    &\times{\Lambda}^{(\textsc{a})}_{nm}(t,\bm x_\textsc{a}){\Lambda}^{(\textsc{b})}_{n'm'}(t',\bm x_\textsc{b})G_R(t,\bm x_\textsc{a};t',\bm x_\textsc{b})\nonumber\\
    &\:\:\:+\hat{\mu}_\textsc{b}(t)\hat{\mu}_\textsc{a}(t')e^{\ii(E^{(\textsc{b})}_{n'm'}t+E^{(\textsc{a})}_{nm}t')}\nonumber\\
    &\:\:\:\:\:\:\times{\Lambda}^{(\textsc{b})}_{n'm'}(t,\bm x_\textsc{b}) {\Lambda}^{(\textsc{a})}_{nm}(t,\bm x_\textsc{a})G_R(t,\bm x_\textsc{b};t',\bm x_\textsc{a}) \Big),
\end{align}
where $E^{(\textsc{i})}_{nm} = \mathcal{E}_n^{(\textsc{i})} - \mathcal{E}_m^{(\textsc{i})}$ and we defined the spacetime smearing functions associated with energy levels $(n,m)$ for system $\tc{I}$ as
\begin{equation}
    {\Lambda}^{(\textsc{i})}_{nm}(t,\bm x) = \chi_\textsc{i}(t)\psi^{(\textsc{i})*}_{n}(\bm x)\psi^{(\textsc{i})}_{m}(\bm x)f_\textsc{i}(\bm{x}).
\end{equation}
In order to obtain the smeared version of the model presented in Subsection \ref{sub:classPointlike}, one must restrict the Hamiltonian in Eq. \eqref{eq:fullClassH} to a single one dimensional eigenspace of the Hamiltonian $\hat{H}_\textsc{i}^{\bm x}$. Physically, this restriction is reasonable if one assumes that the energy gap between the eigenstates of the Hamiltonians $\hat{H}_\textsc{i}^{\bm x}$ is much larger than any other scale involved in the setup. Under this assumption, the transition probability ${|n^{(\textsc{i})}}\rangle\longmapsto|{m^{(\textsc{i})}}\rangle$ is negligible, and the dynamics of the problem can be approximated to only be in the $\mathbb{C}^2$ portion of the Hilbert space $\mathcal{H}$. Restricting the system to a given energy gap then yields the effective interaction Hamiltonian of Eq. \eqref{eq:classHsmearedfinal}.
}

\onecolumngrid

\section{\tbb{Illustration of the interaction of quantum systems via quantum-controlled interactions and interactions with a quantum field}}\label{app:figures}

\begin{figure}[h!]
    \centering
    \includegraphics[width=0.33\textwidth]{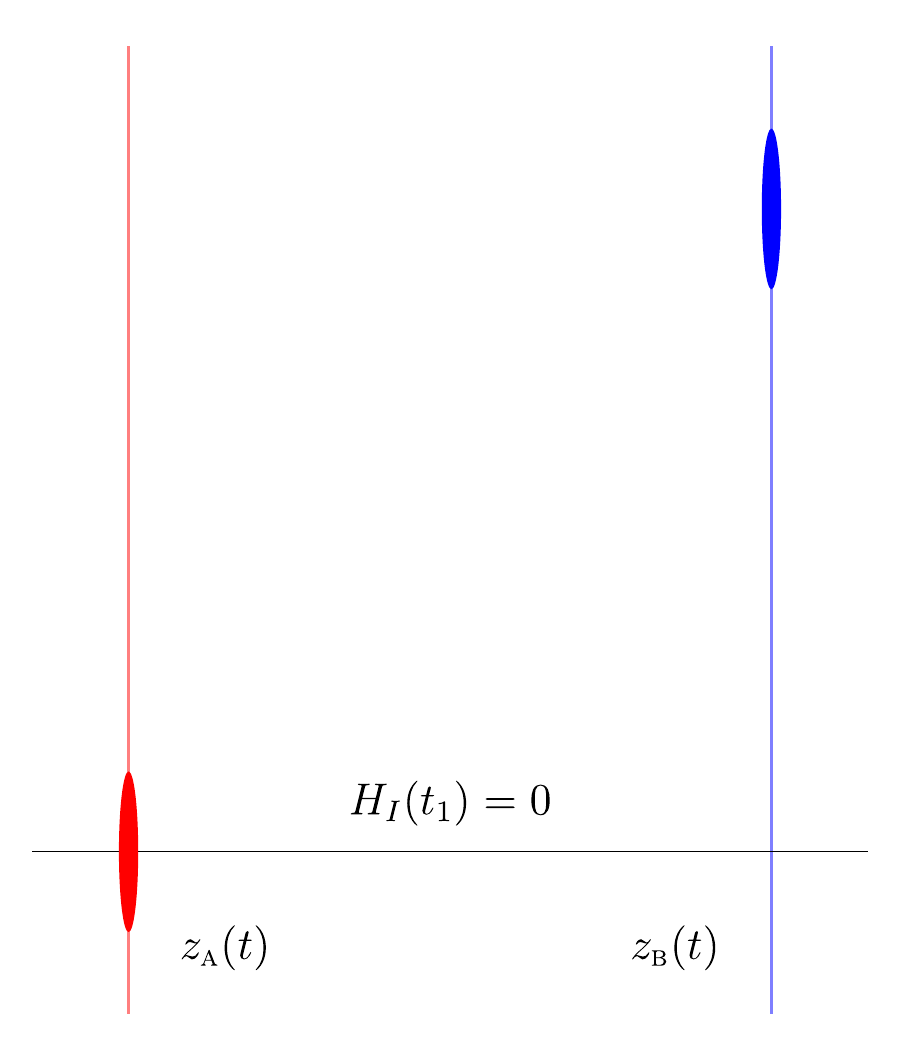}\includegraphics[width=0.33\textwidth]{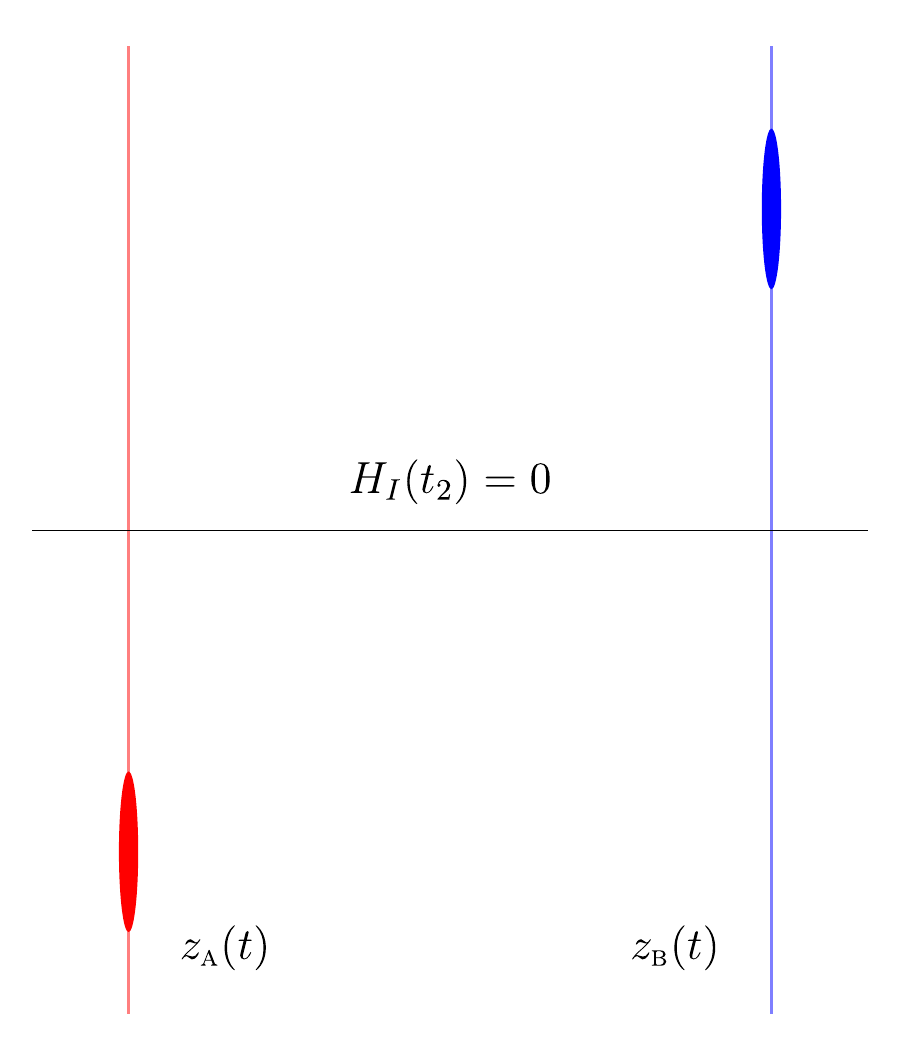}\includegraphics[width=0.33\textwidth]{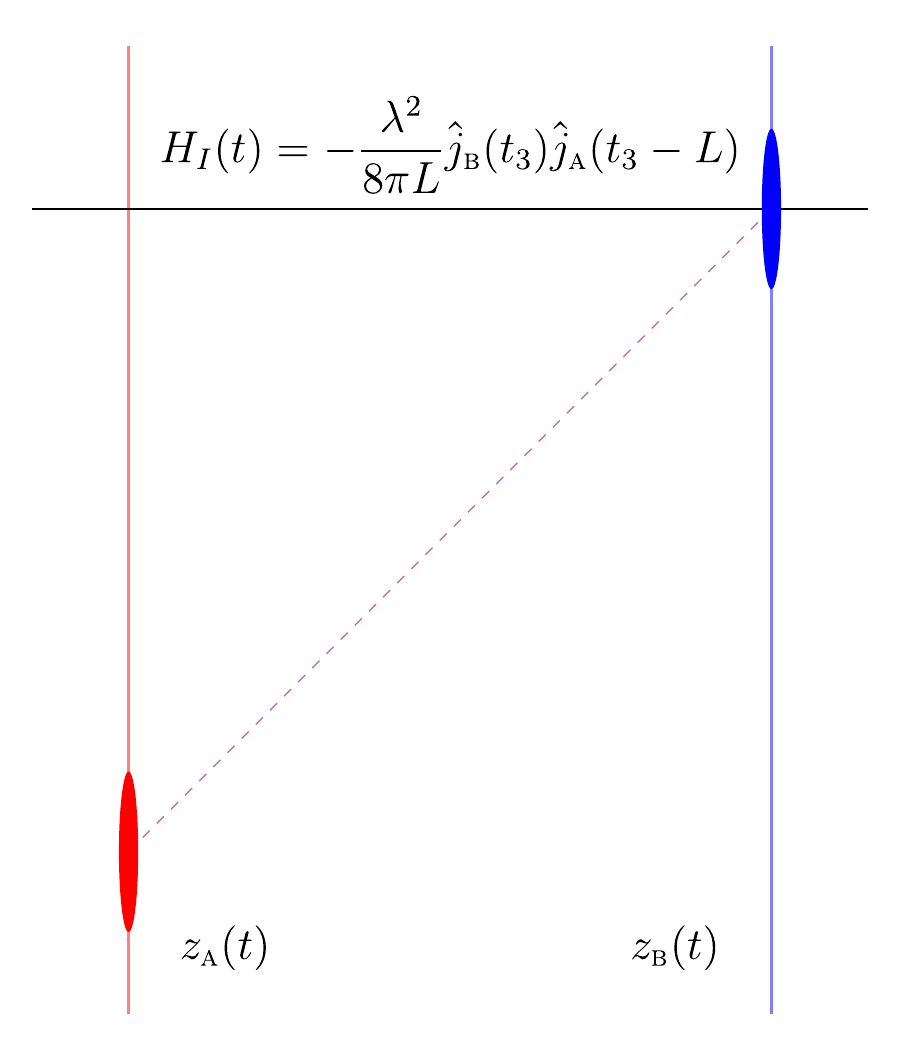}    
    \caption{\tbb{Pictorial representation of the quantum-controlled model mediating the interaction between systems A (red) and B (blue) when they are lightlike separated. Notice that the interaction Hamiltonian only becomes non-zero where the region of interaction of detector A is in the past lightcone of the interaction region of detector B, where the current that sources A appears evaluated at the retarded time $t_3 - L$ due to the retarded Green's function, where $L$ is the separation between the detectors' trajectories in their comoving frame.}}
    \label{fig:qcL}
\end{figure}

\begin{figure}[h!]
    \centering
    \includegraphics[width=0.33\textwidth]{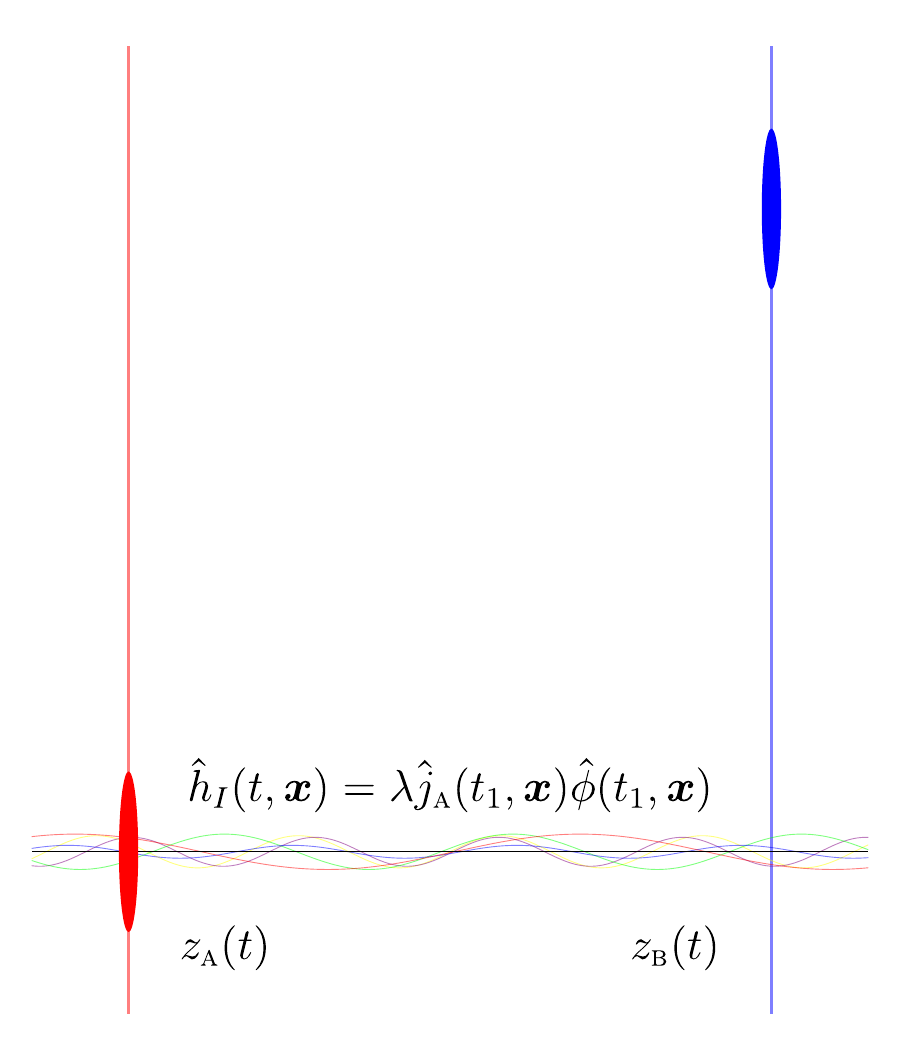}\includegraphics[width=0.33\textwidth]{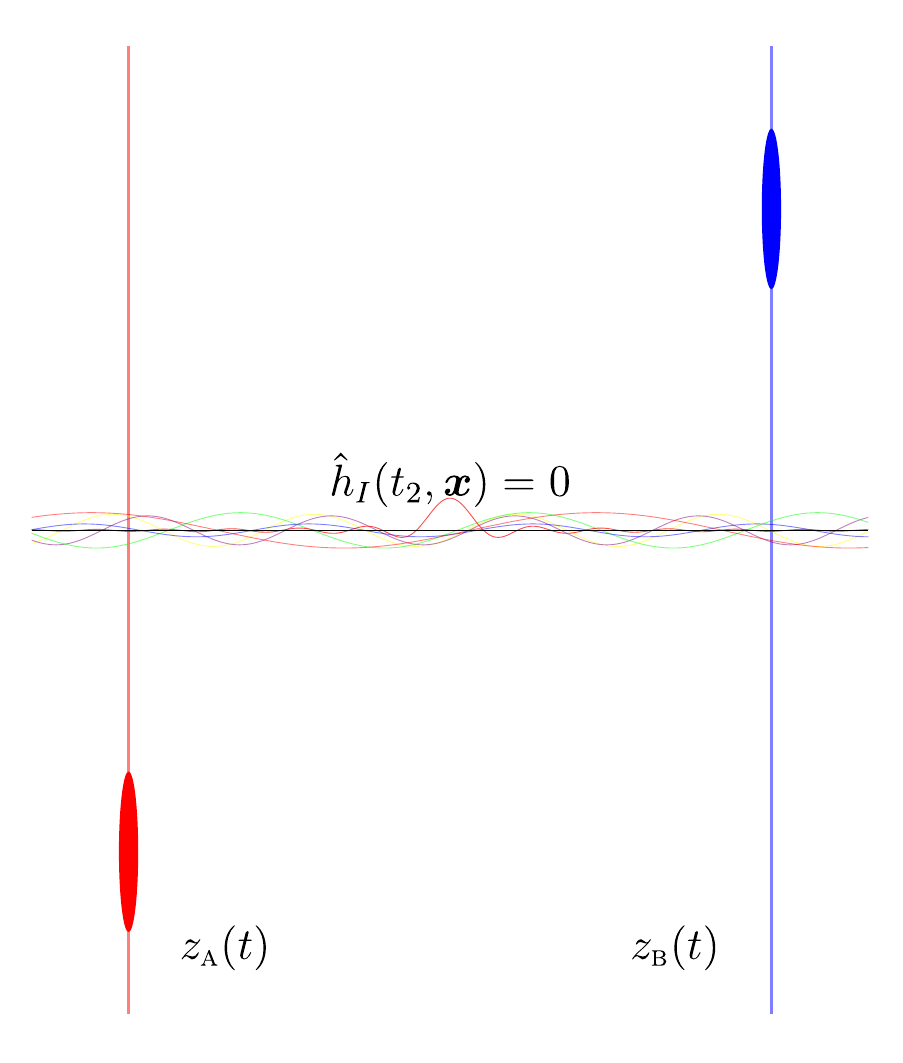}
    \includegraphics[width=0.33\textwidth]{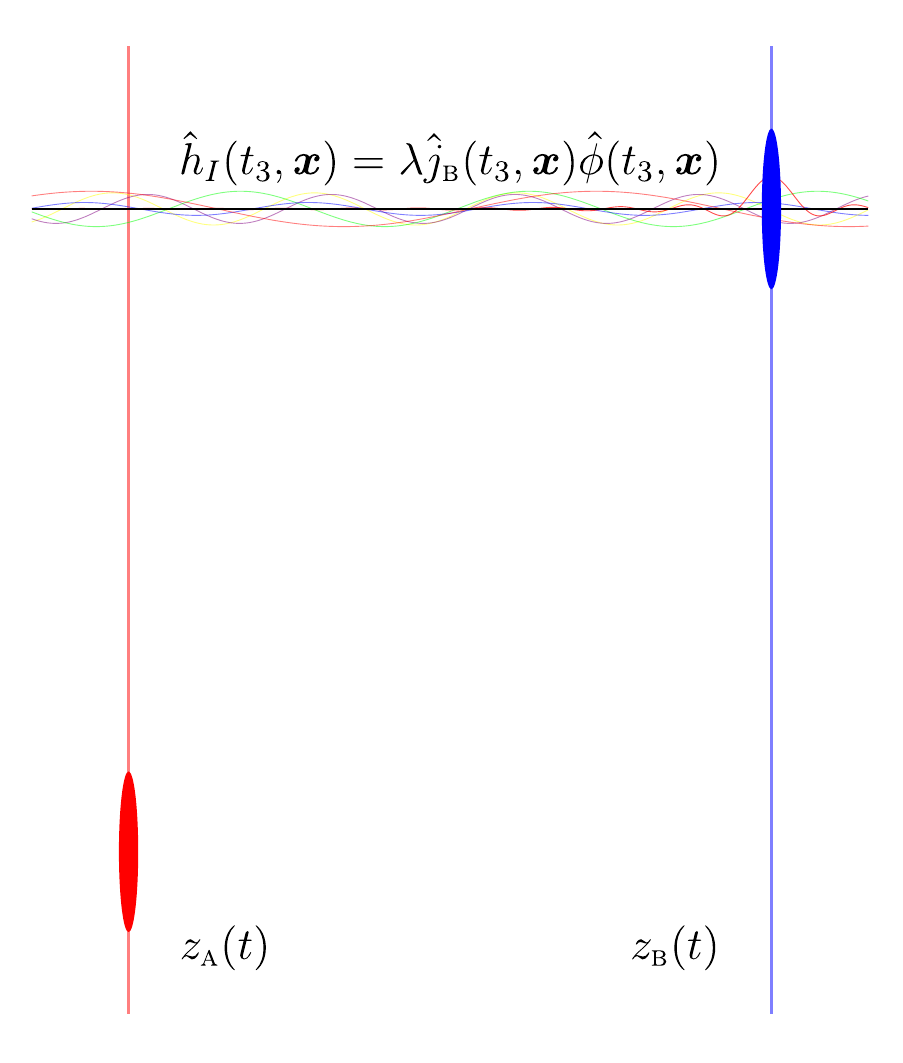}
    \caption{\tbb{Pictorial representation of the quantum field model mediating the interaction between systems A (red) and B (blue) when they are lightlike separated. Notice that the interaction Hamiltonian is non-zero both a the times where detector A interacts with the field (when the correlations of the field become changed due to this interaction) and when detector B interacts with the field (when it probes the field, and may notice the change in correlations due to the interaction of detector A). }}
    \label{fig:qL}
\end{figure}

\begin{figure}[h!]
    \centering
    \includegraphics[width=0.33\textwidth]{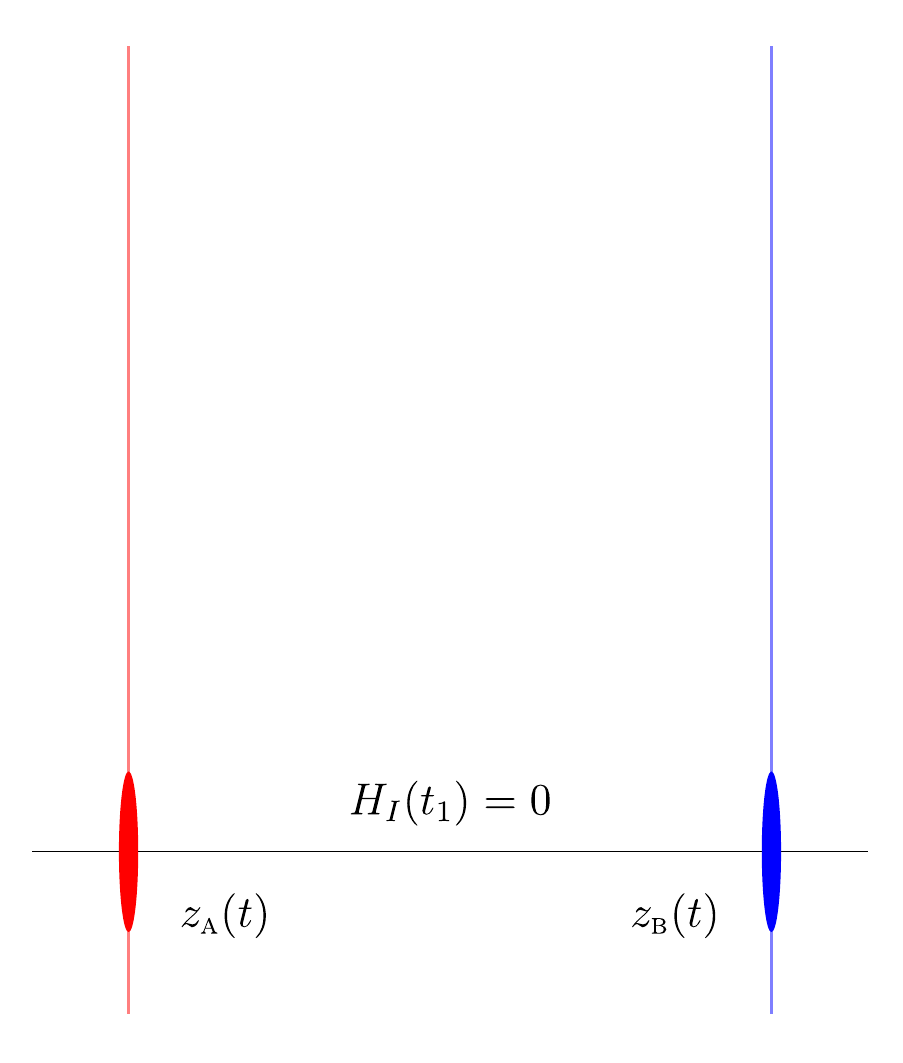}\includegraphics[width=0.33\textwidth]{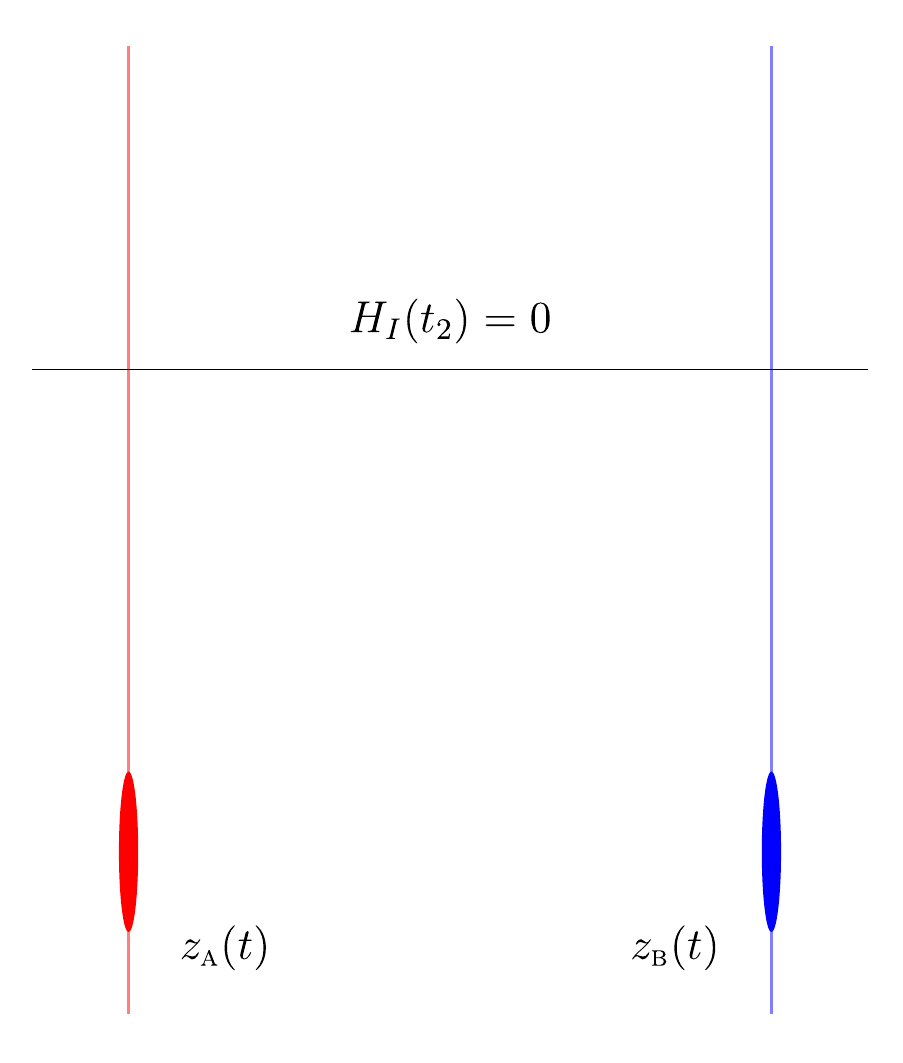}
    \caption{\tbb{Pictorial representation of the quantum -controlled model mediating the interaction between systems A (red) and B (blue) when they are spacelike separated. Notice that the interaction Hamiltonian is always zero, as there is never a time at which the interaction of the detectors is causally connected.}}\label{fig:qcS}
\end{figure}

\begin{figure}[h!]
    \centering
    \includegraphics[width=0.33\textwidth]{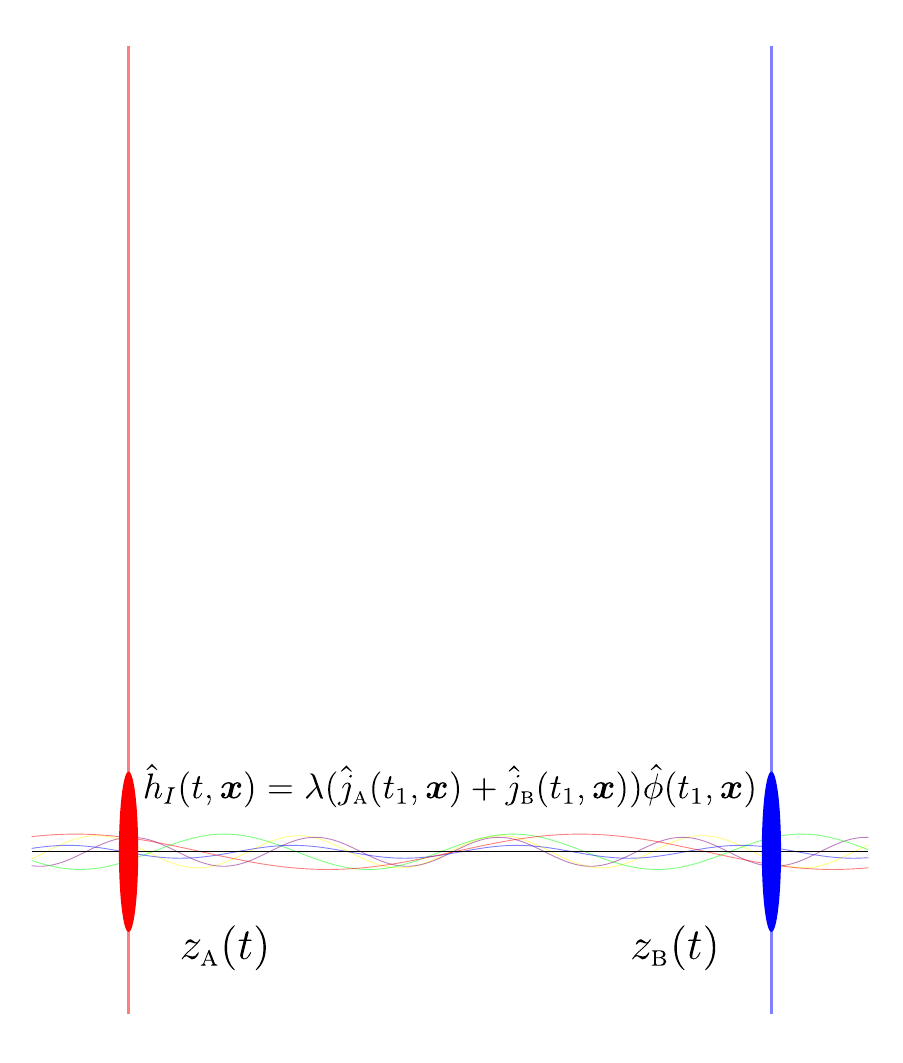}\includegraphics[width=0.33\textwidth]{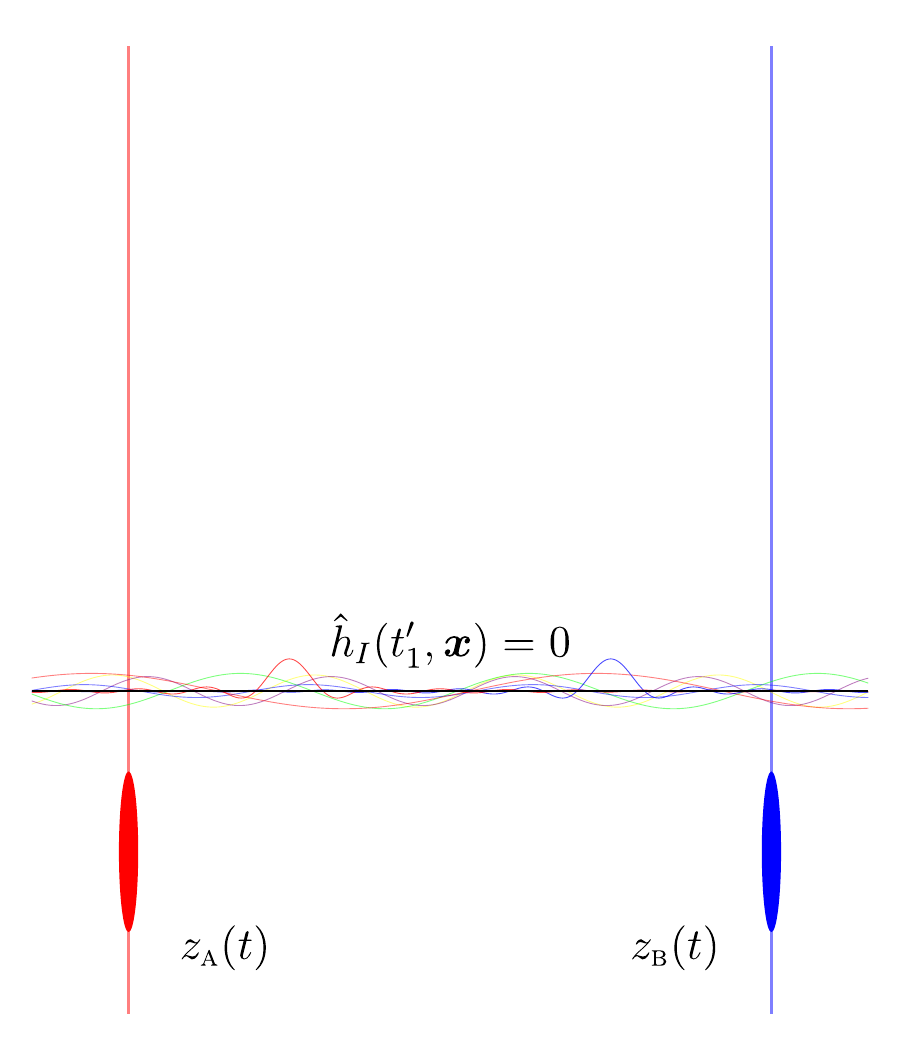}\includegraphics[width=0.33\textwidth]{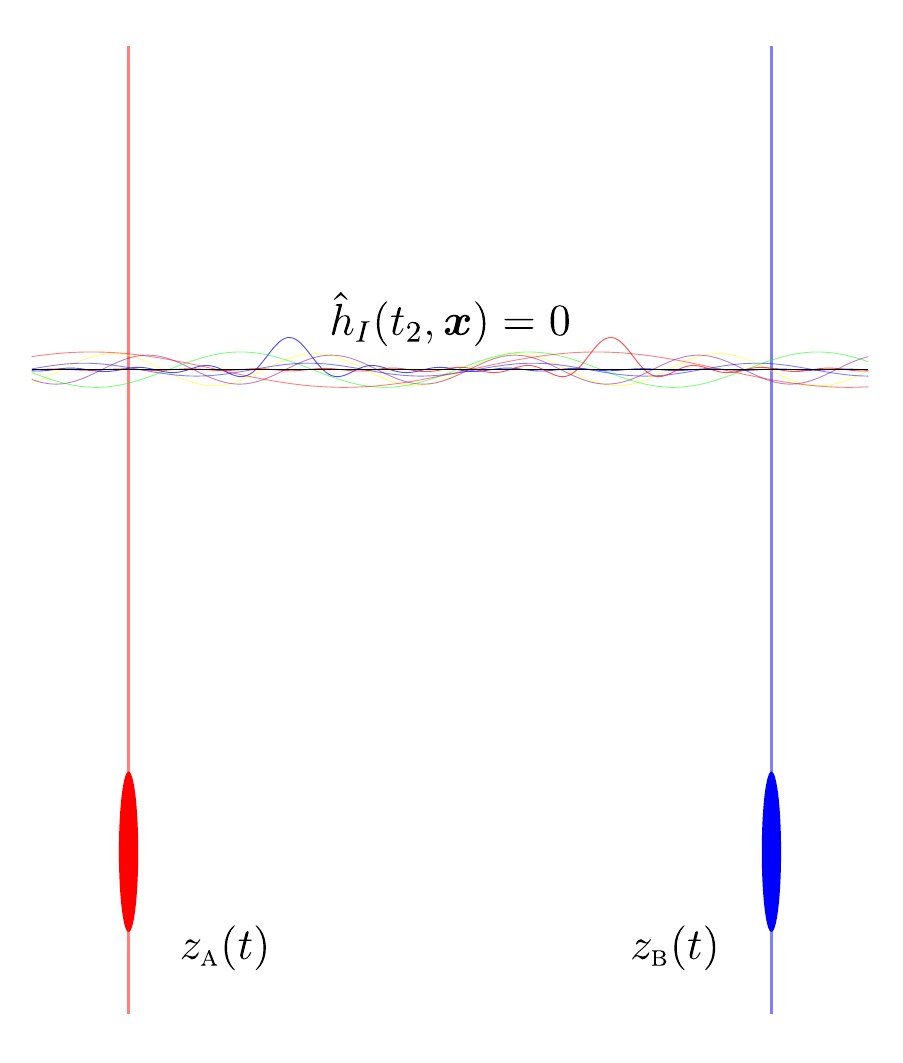}
    \caption{\tbb{Pictorial representation of the quantum field model mediating the interaction between systems A (red) and B (blue) when they are spacelike separated. Notice that the interaction Hamiltonian is non-zero while A and B are interacting with the field. The preexisting entanglement in the field state may then allow the detectors to become entangled with each other through the interaction.}}
    \label{fig:qS}
\end{figure}

\tbb{In this appendix, we present a series of pictorial representations of localized quantum systems that interact via a quantum-controlled fields and with a quantum field when interacting with a massless scalar field. In Figs. \ref{fig:qcL} and \ref{fig:qL} we display the interaction of quantum systems when their interaction regions are lightlike separated for the quantum-controlled and fully quantum cases, respectively. In Figs. \ref{fig:qcS} and \ref{fig:qS} we display the interaction of quantum systems when their interaction regions are spacelike separated for the quantum-controlled and fully quantum cases, respectively.}

\newpage

\newpage

\twocolumngrid

\bibliography{references}
    
\end{document}